\documentclass[prb,twocolumn,showpacs,preprintnumbers,amsmath,amssymb]{revtex4-1}

\usepackage{graphicx,color}
\usepackage{bm}
\usepackage{soul}
\usepackage{dcolumn}
\usepackage{lipsum}
\usepackage{epstopdf}
\usepackage{ulem}

\begin{document}

\preprint{APS/123-QED}

\title{Topologically-driven three-spin chiral exchange
  interactions\\ treated from first principles}

\author{Sergiy Mankovsky}
\affiliation{%
Department of Chemistry/Phys.\ Chemistry, LMU Munich,
Butenandtstrasse 11, D-81377 Munich, Germany \\
}%
\author{Svitlana Polesya}
\affiliation{%
Department of Chemistry/Phys.\ Chemistry, LMU Munich,
Butenandtstrasse 11, D-81377 Munich, Germany \\
}%
\author{Hubert Ebert}
\affiliation{%
Department of Chemistry/Phys.\ Chemistry, LMU Munich,
Butenandtstrasse 11, D-81377 Munich, Germany \\
}%

\date{\today}

\begin{abstract}

  The mechanism behind the three-spin chiral interaction (TCI)
  included in the extended Heisenberg Hamiltonian and represented by an
  expression worked out recently 
(Phys.\ Rev.\ B, {\bf 101}, 174401 (2020))
   is discussed.  
   It is stressed that this approach provides a unique set of the
  multispin exchange parameters which are independent of each other
  either due to their different order of perturbation or due to
  different symmetry.  
  This ensures in particular the specific properties of the TCI
   that were demonstrated previously via fully
   relativistic first principles calculations, and that result from the common
   influence of several issues not explicitly seen from the expression
   for the TCI parameters. Therefore, an
   interpretation of the TCI is suggested, showing explicitly its
  dependence on the relativistic spin-orbit coupling and on the
  topological orbital susceptibility (TOS).
  This is based on
  an expression for the TOS that
  is worked out on the same footing as the expression for the TCI. Using
  first-principles calculations we  demonstrate 
  in addition numerically the common topological
  properties of the TCI and TOS. 
 To demonstrate the role of the 
 relativistic spin-orbit
    coupling (SOC) for the TCI, a so-called 'topological' spin
    susceptibility (TSS) is introduced. This quantity characterizes the
    SOC induced spin magnetic moment on the atom in the presence of
    non-collinear magnetic structure, giving a connection between the
    TOS and TCI. Numerical results again support
   our conclusions.
\end{abstract}

\maketitle

\section{Introduction}

\subsection{Interatomic exchange interaction parameters and their calculation}

  The classical Heisenberg model for interacting spins is a 
  powerful platform used for the investigation of magnetic properties of
  materials,
  taking into account only two-sites bilinear exchange interactions.
  Various schemes have been developed to calculate the corresponding
  model parameters on a first 
  principles level. Most of these are based on the evaluation of the
  energy change due to a distortion of the magnetic subsystem, caused by
  a tilting of the magnetic moments with respect to their orientation
  for a suitable reference spin configuration, which can be chosen either  collinear
  or non-collinear. It should be noted that such an approach, in contrast to 
  model calculations, does no rely on a certain specific exchange
  mechanism and is therefore in principle applicable to any type of
  systems. If there are contributions from
  different types of exchange mechanism, additional
  investigations may be required to clarify the details concerning the
  origin of the exchange interaction (see, e.g. Refs.\
  [\onlinecite{WMW10}] and [\onlinecite{KCS+16}]).

  Many calculations of the exchange parameters reported in the
  literature rely on the idea of the Connolly-Williams (CW)
  method \cite{CW83}. Using the parametrized 
  form of the energy in the Heisenberg model, exchange parameters are
  evaluated within this approach by fitting them to the total energy calculated from first
  principles for different spin configurations \cite{DF04,ALU+08}.
   An alternative scheme applied to 
   calculate the exchange parameters in the momentum space relies
  on the energies corresponding to spin-spiral structures characterized by specific
  wave vectors \cite{USK94,HEPO98,SB02,HBB08}. 
  In contrast to this, one may calculate
  the interatomic exchange interaction parameters directly by 
  evaluating the energy change due to a change of the relative
  orientation of the magnetic moments on two atoms. Such a scheme 
  has been implemented using the Green function (GF) formalism in
  combination with the multiple scattering (KKR) as well as LMTO band
  structure methods \cite{LKG84,LKAG87,PKT+00,USPW03,EM09}. 

  Despite the obvious success of the classical Heisenberg model for many
  applications, it fails to describe more subtle properties of magnetic
  materials without the extension of the Heisenberg Hamiltonian accounting
  among others for higher-order multi-site terms
  \cite{Kit60,KMS+96,MKS97,PP04,GT09,GST14,BCK+14,FESS15,KAE+20}.  
  Similar to the bilinear interaction term, the parameters of the
  extended Heisenberg Hamiltonian may be provided by using results of
  electronic structure calculations. So far, only few first-principles
  calculations have been reported in the literature for that.
  For example, the fourth-order interactions (two-site and three-site) for Cr
  trimers \cite{ALU+08} were calculated using a CW-like method, demonstrating
  their significant magnitude compared to bilinear interactions.
  The fourth-order chiral interactions for a deposited Fe atomic chain
  \cite{LRP+19a} were calculated using the energies for different spin
  configurations.
  In Refs.\ [\onlinecite{PHMH20}] and [\onlinecite{GHMH21}] the biquadratic, three-site four
  spin and four-site four spin interaction parameters have been obtained using
  the energies calculated for different spin configurations
  and applying a CW-like method.
  However, this scheme becomes more and more demanding when
    including higher order multispin interaction parameters, with 
  the difficulties caused by the correspondingly increasing number of spin
  configurations required for mapping of their first-principles energies
  to the increasing number of parameters.
  Another, more flexible  mapping scheme using perturbation theory within the KKR
  Green function formalism was only recently reported by
  Brinker et  al. \cite{BSL19,BSL20}, and by the present authors \cite{MPE20}.
  Note that in contrast to the energy fitting scheme for the determination of
  multispin interaction parameters, the latter approach allows a safe extension of
  the series of contributing terms to the spin Hamiltonian, such that
  including higher-order terms has no impact on the lower-order terms.

  Finally, one has to stress that the exchange coupling
  parameters may depend substantially on the chosen reference spin configuration, as
  was shown, e.g., in Refs.\ [\onlinecite{SKMR05}] and [\onlinecite{BSW08}] by comparing the
  bilinear isotropic exchange parameters obtained for disordered local
  moment (DLM) and ferromagnetic (FM) states. 
  This implies that the reference configuration should be close to
  the magnetic state to be described,
  as the corresponding calculated exchange parameters ensure a better
  description of the magnetic properties of the system.
  This allows in particular to minimize the contribution of 
  higher-order terms in the spin Hamiltonian.  
  A technique developed for the calculation of the exchange
  parameters for a non-collinear reference state has been reported by
  Szilva et al. \cite{STB+17,SCB+20}.
   The resulting spin Hamiltonian obtained on the basis of a 'predefined
   spin configuration', is classified
   as local Hamiltonian by these authors \cite{SSB+21} and is expected to need
   no further multispin expansion as the bilinear terms
   should account for these contributions.
   The authors demonstrate, that within this approach a term similar to
   the Dzyaloshinskii-Moriya interaction (DMI) term may occur even in cases when
  the standard prerequisites for the occurrence of the DMI in case of
  collinear magnetic systems, i.e. spin-orbit coupling (SOC) and  
  lack of inversion symmetry, are not given \cite{CBS+20,SCB+20}. This
  finding has been associated with the contributions of multispin interactions
  incorporated in this  DMI-like term. 
  However, as a restriction intrinsic for this approach
    one has to keep in
  mind, that the parameters calculated for a 'predefined spin
  configuration' are reliable only in the vicinity of this configuration. 
  
  The approaches mentioned so far are of restricted use for an energy
  mapping when the system is brought out of
  equilibrium, e.g., by a strong and ultrafast laser pulse, as this
  situation needs a calculation of the exchange parameters beyond the
  adiabatic approximation. A corresponding theoretical formalism based on 
  non-equilibrium Green functions developed by Secchi et
  al. \cite{SBLK13} gives access only to pair dynamical exchange
  parameters that are expected to represent as well
  contributions of multispin interactions. Similar to
  the findings for the non-collinear reference state
  \cite{CBS+20,SCB+20}, the authors report about the occurrence of a
  DMI-like exchange coupling, called 'twist exchange' that has a
  non-relativistic origin and is 
  attributed by the authors to higher-order three-spin interactions.

  Formally, the multisite expansion includes terms characterizing
  the interaction of any number of spin moments.
  Among these interactions, the three-site three-spin chiral
  interactions (TCI) attracted special attention. On the one hand side,
  TCI can play a crucial role for the proprieties of chiral spin
  liquids, as it was discussed in the literature \cite{GT09,GST14,BCK+14}.
  On the other side, the occurrence of this term (as well as others
  $(2n+1)$-spin interactions) have recently been questioned \cite{HB20,SBL+21a},
  because of the time-inversion asymmetry of the three-spin interaction.
  The explicit calculation in our previous work \cite{MPE20} of the TCI
  parameter and its properties, in particular with respect to 
  time reversal, show that the energy contribution due to the TCI is
  invariant w.r.t. time reversal, and for this reason should
  therefore be considered in the spin Hamiltonian. 
 Another origin for a three-spin chiral interaction
 stem from four-spin interactions, as suggested  
 by Grytsiuk et al.\  \cite{GHH+20}, stressing that
 such a term may give rise for a three-spin interaction invariant
 with respect  to time reversal. 
In line with this, dos Santos Dias et al.\ \cite{SBL+21a}
 considered the three-spin chiral interactions 
 treated as a particular case
  of the 'proper chiral four-spin interaction' 
  as worked out in  Ref.\ [\onlinecite{GHH+20}].
Based on their work Santos Dias et al.\ conclude that the TCI 
discussed in Ref.\ [\onlinecite{MPE20}] seems to be
misinterpreted in spite of the numerical results 
presented in Ref.\ [\onlinecite{MPE20}] that are
 not doubted by these authors.
  Among others, our presentation below demonstrates the misleading
   character of their arguments.

\subsection{Multi-site expansion of spin Hamiltonian: General remarks}

Discussing the multi-site extension of the Heisenberg Hamiltonian in
our recent work \cite{MPE20}, the total energy calculated from first
principles is mapped  onto the spin Hamiltonian 
\begin{eqnarray}
  H &=&  - \sum_{i,j}  J^{s}_{ij} (\hat{s}_i \cdot \hat{s}_j)
 - \sum_{i,j}  \vec{{D}}_{ij} \cdot (\hat{s}_i \times
      \hat{s}_j) \nonumber \\ 
 &&    
 - \frac{1}{3!}\sum_{i,j,k}
                   J_{ijk} \hat{s}_i\cdot (\hat{s}_j \times \hat{s}_k) \; ,  \nonumber \\          
 &&      -  \frac{2}{p!}\sum_{i,j,k,l}  J^{s}_{ijkl} (\hat{s}_i \cdot
        \hat{s}_j)(\hat{s}_k \cdot \hat{s}_l)  \nonumber \\  
 &&  
 - \frac{2}{p!} \sum_{i,j,k,l}  \vec{{\cal D}}_{ijkl} \cdot (\hat{s}_i \times
      \hat{s}_j) (\hat{s}_k \cdot \hat{s}_l) + ... \;,
\label{Eq_Heisenberg_general}
\end{eqnarray}
where $p$  specifies the number of interacting atoms or spins, respectively, and
the parameters ($J^{s}_{ij}, \vec{{D}}_{ij}$, etc.) represent the various
types of interatomic interaction \cite{MPE20}.
Note that terms giving rise to magnetic anisotropy are omitted in the Eq.\
(\ref{Eq_Heisenberg_general}), as we are going to discuss only pure 
exchange interaction terms.
 
  As in our previous work we assume that the dependence of the total
  energy on the magnetic configuration is obtained by perturbation
  theory with respect to a suitable reference state.
  In the following we discuss the implication for the mapping on the
  Hamiltonian given in Eq.\ (\ref{Eq_Heisenberg_general}).  

 First, one should note that each interaction term in Eq.\
  (\ref{Eq_Heisenberg_general}) is characterized by its 
intrinsic properties with respect to a permutation of the interacting
spin moments, i.e.\ it is symmetric, antisymmetric or non-defined
w.r.t. such a permutation. This symmetry is determined by the
combination of scalar and vector products of different pairs of spin
moments, occurring in a specific way.

Second, the first-principles expressions for the exchange parameters are
derived in a one-to-one manner following the properties of the spin-products characterizing
different terms of the spin Hamiltonian, ensuring this way unique permutation
properties for the corresponding exchange interaction term. 
Treating the $p$-spin exchange interactions
$\underline{J}_{i_1,i_2,... i_p}$ 
 (where $\underline{J}$ indicates
hidden indices of the tensor $J^{\nu_1 \nu_2
  ... \nu_p}_{i_1,i_2,... i_p}$, $\nu_i = \{x, y, z\}$
 written here as superscripts only for the sake of convenience) 
in terms of the rank-$p$ tensor in the $3p$-dimensional subspace of the
interacting spin moments $\{\hat{s}_{i_1},\hat{s}_{i_1},...,\hat{s}_{i_p}\}$,
this implies a symmetrization of the tensor with respect to the
permutation of a certain set of indices.
Or, the other way around, different types of $p$-spin exchange interactions can be
associated with the tensor forms symmetrized or antisymmetrized with
respect to a permutation of certain indices.
For instance, one can 
distinguish between different symmerized tensor forms \cite{RC05}:
$\underline{J}_{(i,j),(k,l)} = (1/4)(\underline{J}_{i,j,k,l} +
\underline{J}_{j,i,k,l} + \underline{J}_{i,j,l,k} + 
\underline{J}_{j,i,l,k})$, $\underline{J}_{(i,j),[k,l]} =
(1/4)(\underline{J}_{i,j,k,l} + \underline{J}_{j,i,k,l} - \underline{J}_{i,j,l,k} - 
\underline{J}_{j,i,l,k})$, or $\underline{J}_{[i,j],[k,l]} =
(1/4)(\underline{J}_{i,j,k,l} - \underline{J}_{j,i,k,l} -
\underline{J}_{i,j,l,k} + \underline{J}_{j,i,l,k})$ characterizing the 4-spin
interaction terms associated with the  $(\hat{s}_i \cdot
\hat{s}_j)(\hat{s}_k \cdot \hat{s}_l)$, $(\hat{s}_i 
\cdot \hat{s}_j)(\hat{s}_k \times \hat{s}_l)$, and  $(\hat{s}_i \times
\hat{s}_j)(\hat{s}_k \times \hat{s}_l)$ spin products, respectively.
Note that also the shape of each symmetrized element $\underline{J}$ is
determined by the symmetrization with respect to permutations, as it was
demonstrated by Udvardi et al.\ \cite{USPW03} for the DMI, as a
particular case of bilinear interactions.    
The symmetrized interactions can not be transformed one into another as
they correspond to different representations of the permutation
group. In particular, there is no 
connection between the $\underline{J}_{[i,j],[k,l]}$ and
$\underline{J}_{(i,j),(k,l)}$ 
interaction parameters despite Lagrange's identity $(\hat{s}_i \times
\hat{s}_j)(\hat{s}_k \times \hat{s}_l) = (\hat{s}_i \cdot
\hat{s}_k)(\hat{s}_j \cdot \hat{s}_l) - (\hat{s}_i \cdot
\hat{s}_l)(\hat{s}_j \cdot \hat{s}_k)$ that relates
the mixed cross product of the spin moments to a combination of 
scalar products.

Third, each higher-order term in Eq.\ (\ref{Eq_Heisenberg_general})
can be related in a one-to-one manner to a higher-order term of an energy
expansion connected with the perturbation caused by spin tiltings
\cite{LRP+19a,BSL20,GHH+20, MPE20}. 
As a consequence they give an additional energy contribution missing in the
lower-order energy expansion. This implies, carrying the expansion to
higher and higher order does not change the results for the lower-order
terms - in contrast to a fitting procedure.
Moreover, having the same symmetry properties with respect to a permutation
of the indices, the higher-order term can be seen as a
correction to a corresponding lower-order term, e.g. as it takes place
in the case of DMI and 4-spin DMI-like terms. 

 Finally, it should be added that
 one has to distinguish the chiral properties of the
 DMI-like interactions arising from 
local inversion symmetry being absent and the
 topologically-driven chiral properties of the TCI 
considered in Ref.\ \cite{MPE20}. 
 This implies among others 
 that the four-spin chiral interactions discussed in 
 Refs.\ [\onlinecite{LRP+19a}] and [\onlinecite{GHH+20}], have no connection with 
 the TCI worked out in Ref.\ [\onlinecite{MPE20}] and considered here.
Nevertheless, the  SOC plays a central role in
 both cases.

 In the present contribution we are going to discuss in more details
  the origin of the TCI that was derived in
  Ref.\ [\onlinecite{MPE20}] and its specific features in
  comparison with the chiral interactions represented by the expressions
  worked out in Refs.\ [\onlinecite{LRP+19a}] and [\onlinecite{GHH+20}]. 
  For this, we give in Sec.\ \ref{TCI_frel} the expression for TCI
  \cite{MPE20} based on a fully relativistic approach, accompanied with
  some comments concerning its properties. 
In Sec.\ \ref{TCI+SOC} we will show explicitly the role of
 the relativistic spin-orbit interaction for the TCI, which is different
 when compared to its role for the expressions reported in Refs.\
 [\onlinecite{LRP+19a}] and [\onlinecite{GHH+20}]. Moreover, 
 we demonstrate in this section an explicit interconnection of the TCI with the  
 topological orbital susceptibility (TOS) for triples of atoms, which determines
 the chiral properties of the TCI. 
 Details of the properties of the TOS will be discussed in Sec.\ \ref{SEC:TCI-TOM}.
 For this the corresponding expression is derived within the fully
 relativistic approach on the same footing as for the TCI. We will compare the
 results for the topological orbital moment (TOM) calculated by means of
 the TOS 
 with results obtained self-consistently for embedded 3-atomic clusters. 
 To allow a more detailed discussion of the TCI,
we introduce in Sec.\ \ref{TSS} the 'topological' spin susceptibility
(TSS), in analogy to the topological orbital susceptibility. All
discussions and formal developments are accompanied by corresponding
numerical results supporting our conclusions on the TCI's origin.
Some more technical aspects of this work concerning the
properties of the TCI with respect to time reversal,
computational details, and
the expression for the topological orbital moment (TOM)
are dealt with in some detail in three appendices.

\section{Three-spin chiral exchange interactions from first-principles} \label{SEC:TCI}

In the following, we discuss some specific properties of the three-spin chiral
interaction (TCI) term in the spin Hamiltonian, which ensure that it
cannot be represented in terms of interactions having different
permutation properties.

\subsection{TCI via fully relativistic
  approach \label{TCI_frel}}

Focusing on the properties of the TCI, we give here the expression
derived within the multiple-scattering formalism 
\cite{MPE20}
%
\begin{eqnarray}
  J_{ijk}  &=&   \frac {1}{4\pi} \mbox{Im}\, \mbox{Tr} \int^{E_F} dE (E - E_F)\,  \nonumber \\
 &&                    
\times \Big[ \underline{T}^{i, x}\, \underline{\tau}^{ij}
\underline{T}^{j, y}\, \underline{\tau}^{jk}
               \underline{O}^{k}\,  \underline{\tau}^{ki} 
- \underline{T}^{i, y}\, \underline{\tau}^{ij}
 \underline{T}^{j, x}\, \underline{\tau}^{jk}
     \underline{O}^{k}\,  \underline{\tau}^{ki}
               \nonumber \\       
&&     - \underline{T}^{i, x}\, \underline{\tau}^{ij}
\underline{O}^{j}\, \underline{\tau}^{jk}
               \underline{T}^{k, y}\, \underline{\tau}^{ki}\;
+ \underline{T}^{i, y}\, \underline{\tau}^{ij}
\underline{O}^{j}\, \underline{\tau}^{jk}
               \underline{T}^{k, x}\, \underline{\tau}^{ki} \;
               \nonumber \\       
&& +
   \underline{O}^{i} \, \underline{\tau}^{ij} 
\underline{T}^{j, x}\, \underline{\tau}^{jk}
               \underline{T}^{k, y}\, \underline{\tau}^{ki}\;
 - \underline{O}^{i}\,  \underline{\tau}^{ij}
\underline{T}^{j, y}\,\underline{\tau}^{jk}
               \underline{T}^{k, x}\, \underline{\tau}^{ki} \Big] \;,
               \nonumber \\       
\label{Eq:J_XYZ} 
\end{eqnarray}
where the matrix elements of the torque operator
$T^{i,\alpha}_{\Lambda\Lambda'}$ and the overlap integrals $O^{i}_{\Lambda\Lambda'}$ 
are defined as follows:\cite{EM09a}
%
\begin{eqnarray}
 T^{i,\alpha}_{\Lambda\Lambda'} & = & \int_{V_i} d^3r  \, Z^{i \times}_{\Lambda}(\vec{r},E)\, \Big[\beta \sigma_{\alpha} B_{xc}^i(\vec{r})\Big] \, Z^{i}_{\Lambda'}(\vec{r},E)\;.  \label{Eq:MET}
\end{eqnarray}
and 
%
\begin{eqnarray}
 O^{i}_{\Lambda\Lambda'} & = & \int_{V_i} d^3r  \,
 Z^{i \times}_{\Lambda}(\vec{r},E) \, Z^{i}_{\Lambda'}(\vec{r},E).  \label{Eq:MEO}
\end{eqnarray}
Here ${\vec B}_{xc}(\vec r)$ is the spin-dependent part of the 
exchange-correlation potential,
 $\vec{\sigma}$  is the vector of $4 \times 4$ Pauli matrices and 
$\beta$ is one of the  standard Dirac matrices \cite{Ros61,EBKM16};
$Z^{n}_{\Lambda_1}(\vec{r},E)$ and 
$J^{n}_{\Lambda_1}(\vec{r},E)$ 
are
the regular and irregular  solutions of the
single site Dirac equation and
  ${\underline{\tau}}^{n n'}$ is the scattering path operator matrix \cite{EBKM16}.

As one can see, the 
expression in Eq.\ (\ref{Eq:J_XYZ}) ensures the properties specific only for
the TCI. I.e.\ (i) the anti-symmetry with respect to permutation
  of any two spin indices, and (ii) the invariance with respect to
a cyclic permutation of the spin indices
in the $i,j,k$ sequence, that is the result of the 
invariance of the trace upon cyclic permutation of the product of matrices.
In addition, it was shown in our previous work \cite{MPE20} that the TCI
parameter is antisymmetric with respect to time reversal, leading to
an invariance with respect to time reversal for the energy contribution
associated with this interaction
(see also Appendix \ref{SEC:TR}).

  Obviously, an energy expansion 
to higher orders, as indicated in 
Eq.\ (\ref{Eq_Heisenberg_general}),   
  includes more interaction
  terms that are anti-symmetric with respect to permutations of the three
  indices $i, j, k$, giving the energy contribution $\sim J_{(i,[j,k]),l...,n}
  (\hat{s}_i\cdot[\hat{s}_j\times \hat{s}_k]) ... $. These contributions, however, 
  have a more complicated dependence on the magnetic
  configuration when compared to the TCI as
   noted already in Ref.\  [\onlinecite{MPE20}], 
   because more spins are involved in the interaction.
  Moreover, in accordance with the discussion above, one has to stress
  once again, that the TCI is
  uniquely determined by symmetry and cannot be represented in
  terms of other interactions that have 
 higher order  or 
  different symmetry with respect to 
  a permutation of the spin indices.
  This implies  in particular, an independence on the parameters $\sim
  J_{(i,[j,k]),l}$ and those given in the previous section, when
  discussing the 4th-order interactions.

\subsection{TCI and relativistic spin-orbit coupling in
  terms of non-relativistic Green functions  \label{TCI+SOC}}

 In this section we discuss the mechanism responsible for
  the TCI reported in Ref.\ [\onlinecite{MPE20}], to distinguish it from
  the fourth-order three-spin interactions suggested  by Grytsiuk et al.\
  \cite{GHH+20}.  
  Using the nonrelativistic Green-function-based description, one can
  factorize the expression for the TCI, to represent it in
  terms of the relativistic SOC and the topological orbital susceptibility (TOS).
  This factorization allows us to show explicitly the role of the
  SOC for the mechanism leading to the TCI \cite{MPE20}, and to
  demonstrate that it is different when compared to the mechanism
  responsible for three-spin chiral interaction discussed in Refs.\
  [\onlinecite{GHH+20}] and [\onlinecite{SBL+21a}]. 
%
%

  According to the latter work,  
the three-spin chiral interaction is associated with
a topological orbital moment $\vec{L}^{\rm{TO}}_{ijk}$ induced on the atoms
of every triangle formed by magnetic atoms, $\Delta_{ijk}$, due to the
non-coplanar orientation of their spin magnetic moments.
As it was suggested for the case of all  atoms being equivalent, the TCI
term can be written as $~\sim \xi \chi^{\rm{TO}}_{ijk}
\hat{s}_i\cdot(\hat{s}_j\times \hat{s}_k) (\hat{n}_{ijk}\cdot \langle 
\hat{s} \rangle )$, where $\langle \hat{s} \rangle = \frac{1}{3}(\hat{s}_i +
\hat{s}_j + \hat{s}_k)$, $\chi^{\rm{TO}}_{ijk}$ is the topological
orbital susceptibility and $\xi$ is the relativistic spin-orbit
interaction parameter corresponding to atom $i$ with spin moment
$\vec{s}_i$.  This expression shows explicitly the dependence 
of the three-spin interaction on the orientation of a sum of the
interacting spin magnetic moments with respect to the normal vector
$\hat{n}_{ijk}$ of a triangle $\Delta_{ijk}$. In particular, it implies
that the TCI is proportional to the flux of the local spin magnetization
through the triangle area. 

 Here we discuss in more details the mechanism giving
  rise to the TCI reported in  Ref.\ [\onlinecite{MPE20}].
As a reference, we start from the ferromagnetic (FM) state of the system
with the magnetization aligned  
along $\hat{z}$ direction, and its electronic structure
characterized by the Green function $G_0$. To demonstrate explicitly
the role of the spin-orbit interaction, we consider the Green function $G_0$ in
the non-relativistic (or scalar-relativistic) approximation. For the FM
state considered, it has  spin-block-diagonal form in the global frame of reference.
Creating a non-coplanar magnetic configuration characterized by a finite
scalar spin chirality, we assume infinitesimal tilting 
angles of the spin moments on the interacting atoms, that allows to use
perturbation theory to describe the Green function $G$ of the system with
tilted spin moments as:
\begin{eqnarray}
G  &=&  G_0 + \Delta {G} \;,
\label{Eq_INTERPR-1}
\end{eqnarray}
 where $ \Delta {G} $
  is induced by the tilting of 
  three spin moments $\hat{s}_i, \hat{s}_j $, and $\hat{s}_k$
  represented by the tilting vectors
 $\delta \hat{s}_{i}, \delta \hat{s}_{j}$, and $ \delta \hat{s}_{k}$, respectively.
As it is already well known, the chiral magnetic structure induces a
persistent electric current in the magnetic system, creating that way a
finite orbital moment in addition to that induced by the relativistic 
spin-orbit coupling (SOC)  \cite{NIF03,FNI04,DBB+16a}. Note that the current can be split into a delocalized
part and one localized on the atoms \cite{Tho11}. For the sake of simplicity, we will
focus on the latter one coupled to the spin degree of freedom of the
electrons responsible for the spin magnetic moments of each atom. 
In this case one can speak about a spin magnetic moment $\delta m$ on
the atoms, induced via SOC by the orbital moment created by the chiral
magnetic structure. The induced spin magnetic moment leads in turn to a change of
the exchange-correlation energy
\begin{eqnarray}
  \Delta E_{xc} &=& \int d^3r \frac{\partial E_{xc}[n,m]}{\partial
                    \vec{m}} \cdot \delta \vec{m}(\vec{r}) = -\int d^3r \vec{B}_{xc}(\vec{r})
                    \cdot \delta \vec{m}(\vec{r}) \nonumber \\
\label{Eq_INTERPR-2}
\end{eqnarray}
where $\vec{B}_{xc}(\vec{r}) = \hat{m} B_{xc}[n,m](\vec{r})$ is an effective exchange
field characterizing the spin-dependent part of the exchange-correlation
potential. Here $\hat{m}$ is the direction of the magnetization, and for
the sake of simplicity 
$\vec{B}_{xc}(\vec{r})$ is supposed to be collinear within the cell.

The energy change due to the spin moment induced on the atom $i$
of the considered trimer is given by the integral over its volume $V_i$
\begin{eqnarray}
  \Delta E_{xc,i}(\delta \hat{s}_{i}, \delta \hat{s}_{j},
  \delta \hat{s}_{k}) &=& -\int_{V_i} d^3r B_{xc}(\vec{r}) \, \hat{m} \cdot \delta \vec{m}_i(\vec{r})      
\label{Eq_INTERPR-3}
\end{eqnarray}
with  $\vec{r} \in V_i$ and
\begin{eqnarray}
 && \delta \vec{m}_i(\vec{r}, \delta \hat{s}_{i}, \delta \hat{s}_{j},
  \delta \hat{s}_{k})
   \nonumber \\
  &=& -\frac{1}{\pi} \mbox{Im}\,\mbox{Tr} \int^{E_F} 
                       dE\, \int_{V_{\Delta}} d^3r' \nonumber \\
  && \times \vec{\sigma}\, G_0(\vec{r},\vec{r}\,',E)\, 
     V_{\rm{SOC}}(\vec{r})\, (\vec{\sigma} \cdot \hat{\vec{l}}) \, \Delta G(\vec{r}\,',\vec{r},E) \;,
\label{Eq_T-spin}
\end{eqnarray}
%
where we stress the dependency on the tilting vectors $\delta \hat{s}_{i}, \delta \hat{s}_{j}$,
and $  \delta \hat{s}_{k}$ by including them in the 
argument list.

As we discuss the TCI arising due to the non-coplanar orientation of the
interacting spin moments, the corresponding change of the Green function
can be written as $\Delta G(\vec{r}\,',\vec{r},E) = \Delta G(\vec{r}\,',\vec{r},E,
\delta \hat{s}_{i}, \delta \hat{s}_{j}, \delta \hat{s}_{k}$), for which the
explicit form is discussed in Ref.\ [\onlinecite{MPE20}].
Furthermore, $V_{\Delta}$ in Eq.\ (\ref{Eq_T-spin}) is the volume
corresponding to the interacting atoms $i, j, k$, 
$\hat{\vec{l}}$ is the angular momentum operator and
$V_{\rm{SOC}}(\vec{r}) = \frac{1}{c^2}\frac{1}{r}\frac{\partial
  V(r)}{\partial r}$ for a spherical scalar potential $V(r)$.
This can be rewritten as follows
\begin{eqnarray}
  &&  \Delta E_{xc, i}(\delta \hat{s}_{i}, \delta \hat{s}_{j}, \delta \hat{s}_{k})
   \nonumber \\
  &=& \frac{1}{\pi} \mbox{Im}\,\mbox{Tr} \int^{E_F}
                    dE\, \int_{V_i} d^3r  \int_{V_{\Delta}} d^3r' B_{xc}(\vec{r}) \nonumber \\
  &&  \times (\hat{m} \cdot \vec{\sigma}) G_0(\vec{r},\vec{r}\,',E) \,
                    V_{\rm{SOC}}(\vec{r})\,(\vec{\sigma}\, \cdot \hat{\vec{l}}) \,\Delta
     G(\vec{r}\,',\vec{r},E)   \nonumber \\
                  &=& \frac{1}{\pi} \mbox{Im}\,\mbox{Tr} \int^{E_F}
                    dE\, \int_{V_i} d^3r  \int_{V_{\Delta}} d^3r' B_{xc}(\vec{r}) \nonumber \\
  &&  \times  G_0(\vec{r},\vec{r}\,',E) 
                 \,   V_{\rm{SOC}}(\vec{r})\,(\hat{m} \cdot \hat{\vec{l}})\, \Delta G(\vec{r}\,',\vec{r},E)\,,
\label{Eq_INTERPR-4}
\end{eqnarray}
where we used the expression
\begin{eqnarray}
  \label{eq:pauli1}
  (\vec{\sigma} \cdot \hat{m})(\vec{\sigma}\cdot \hat{\vec{l}})
  &=& \hat{m} \cdot \hat{\vec{l}} + i\vec{\sigma} \cdot (\hat{m} \times
     \hat{\vec{l}}) \,. 
\end{eqnarray}
Taking into account the spin-block-diagonal form
of the non-perturbed Green function, one can show that the second part
of Eq.\ (\ref{eq:pauli1}) can be omitted as the traces (see Ref.\ [\onlinecite{MPE20}])
Tr$(\sigma_{x(y)} \sigma_{x} \sigma_{y})$ and Tr$(\sigma_{x(y)} \sigma_{y}
\sigma_{x})$ are equal to zero. 
 The behavior under time reversal of the energy change in Eq.\
  (\ref{Eq_INTERPR-4}) is discussed in detail in Appendix \ref{SEC:TR}.

As we discuss the three-spin interaction, it is determined by the
chirality-induced energy change according to Eq.\ (\ref{Eq_INTERPR-4}),
i.e.\ $J_{ijk} \sim \Delta E_{xc, i}(\delta \hat{s}_{i}, \delta \hat{s}_{j}, \delta \hat{s}_{k})$.
Moreover, the calculations of the exchange parameters are performed assuming
infinitesimal tilting of the spin magnetic moments in every trimer, that
implies the same orientation $\hat{s}_i = \hat{s}_j = \hat{s}_k =
\hat{m}$ for the reference FM configuration, which gives the dependence
of the three-spin interactions on the orientation of the magnetization
with respect to the surface normal vector $\hat{n}$ of the triangular area. 
Or, the other way around, 
 this implies also that the angle-dependent behavior of the TCI is fully
  determined by the projection of the topological orbital moment (TOM)
  (i.e.\ for vanishing SOC) onto the direction of the magnetization.
  This will be shown below by calculating the orbital moment along the
  magnetization direction oriented along the $z$ axis, for the lattice
 and the normal vector $\hat{n}$ rotated by an angle $\gamma$ within the
 plane perpendicular to the rotation axis.

\begin{figure}
\includegraphics[width=0.2\textwidth,angle=0,clip]{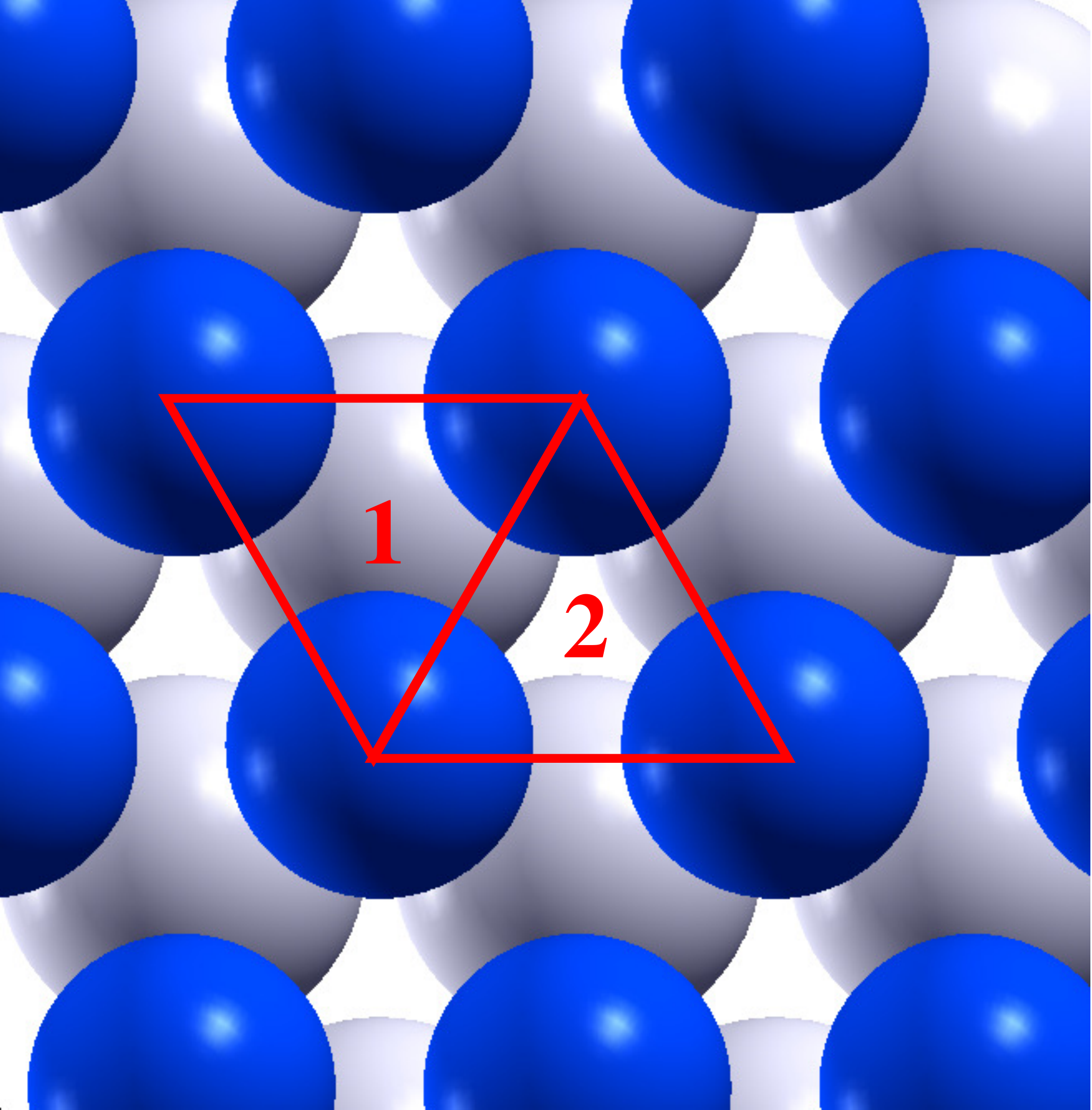}\;
\caption{\label{fig:geometry_triangles} Geometry of the smallest three-atom clusters 
  in the monolayer of 3$d$-atoms on $M(111)$ surface ($M$= Au, Ir):
  $M$-centered triangle $\Delta_1$ and hole-centered triangle $\Delta_2$.     }   
\end{figure}

\begin{figure}
\includegraphics[width=0.4\textwidth,angle=0,clip]{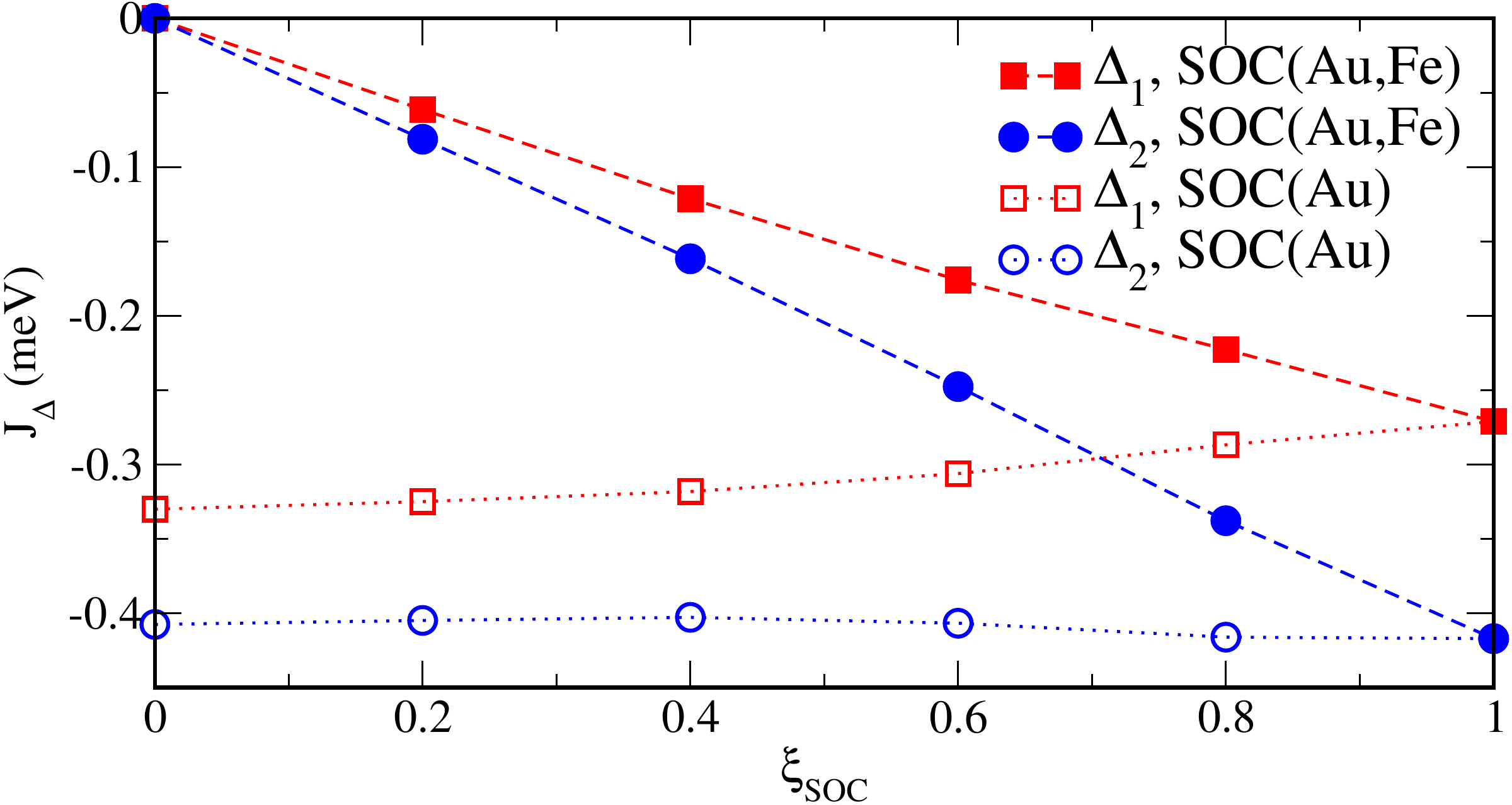}\;
\caption{\label{fig:THEESPIN-SOC} Three-spin chiral exchange interaction (TCI)
  parameters $J_{\Delta}$ calculated for Fe on Au (111) on the basis of
  Eq.\ (\ref{Eq:J_XYZ}) as a function of SOC scaling parameter
  $\xi_{SOC}$ for  the smallest triangles  $\Delta_1$ and $\Delta_2$.
  Full   symbols represent the results obtained when 
 scaling the SOC for all
  elements in the system, while open symbols show the results 
  when scaling  only the SOC for  Au.
}  
\end{figure}

To demonstrate the dependence of the TCI on the relativistic SOC, corresponding
calculations of $J_{\Delta} = J_{ijk} - J_{ikj}$ (see the definition
below), have been performed for 1ML Fe on Au (111), 
for the two smallest
triangles $\Delta_1$ 
and $\Delta_2$ centered at an Au atom or hole site, respectively
(see Fig.\ \ref{fig:geometry_triangles}
for a presentation of the geometry and
Appendix \ref{SEC:Computational-scheme}
for computational details concerning these calculations). 
%
%
Fig.\  \ref{fig:THEESPIN-SOC} gives the parameters
$J_{\Delta_1}(\xi_{\rm{SOC}})$ and
 $J_{\Delta_2}(\xi_{\rm{SOC}})$  calculated using Eq.\ (\ref{Eq:J_XYZ})
 that was derived 
 within the approach reported in our previous work 
 \cite{MPE20}.
 %
%
 Note that setting the SOC scaling factor $\xi_{\rm{SOC}}=0$ implies a suppression
of the  SOC, while $\xi_{\rm{SOC}} = 1$ corresponds to  the
fully relativistic case. 
As expected from Eq.\ (\ref{Eq_INTERPR-4}), we find
indeed a nearly
linear variation of  $J_{\Delta}(\xi_{\rm{SOC}})$ 
with the SOC scaling parameter $\xi_{\rm{SOC}}$ applied to all elements
in the system, shown in Fig.\ \ref{fig:THEESPIN-SOC} by full symbols. 
This shows in particular that the SOC is an ultimate prerequisite for a
finite $J_{\Delta}$ and with this for  
the occurrence of the TCI.
In addition, open symbols in Fig.\
\ref{fig:THEESPIN-SOC} represent the 
parameters  $J_{\Delta}(\xi_{\rm{SOC}})$
calculated when scaling the  SOC  only for Au. 
In this case, one can see only weak
changes of the TCI, reflecting a minor impact of the SOC of the 
substrate on these interactions, in contrast to the DMI-like
interactions that  normally depend strongly  on the SOC for
the substrate atoms.

\subsection{TCI and topological orbital moment} \label{SEC:TCI-TOM}

 In a next step, we derive an expression for the above
  mentioned TOS on the same footing as for the TCI \cite{MPE20},
  i.e. within a fully relativistic approach using the multiple
  scattering GF formalism. By performing the calculations
  on the basis of the
  expressions derived for the TOS and for the TCI, we will demonstrate
  the common properties of these quantities. On the
  other hand, we will perform complementary calculations for the
  topological orbital moment (TOM) using the self-consistent embedded cluster technique,
  to confirm that the derived expression for the TOS gives indeed rise to the
  chirality-induced orbital moment.
  The comparison of the results 
  confirms in particular the topological origin of the TCI.

Thus, for our purpose we represent the TOM  
as a sum over 
the products of the topological orbital susceptibility
  (TOS) $\chi^{\rm{TO}}_{ijk}$ determined 
  for the triples of atoms $(ijk)$ and the
  corresponding scalar spin chirality $\hat{s}_i\cdot (\hat{s}_j \times
  \hat{s}_k) $, that has to be seen as an effective
  inducing magnetic field: 
%
\begin{eqnarray}
  L^{\rm{TO}}  &=& \frac{1}{3!}\sum_{i \neq j\neq k}
                   \chi^{\rm{TO}}_{ijk} \, [\hat{s}_i\cdot (\hat{s}_j \times
              \hat{s}_k)] \;.
\label{Eq_TOM}
\end{eqnarray}
Here we restrict to the component $L^{\rm{TO}} = L_z^{\rm{TO}}$
along the axis $\hat{z}$  of the global frame of reference  
which is taken parallel to the    magnetization  $\vec{m}$
of the  FM reference  system, 
i.e.\  $ \hat{z} \| \hat{m} $ (see \ref{TCI+SOC}).
As one can see, Eq. (\ref{Eq_TOM}) has 
by construction a form similar to the energy
contribution due to the TCI, i.e.\ the third term in Eq.\
(\ref{Eq_Heisenberg_general}). Therefore we will follow
Ref.\ [\onlinecite{MPE20}] to derive an expression for the TOS
$\chi^{\rm{TO}}_{ijk}$.  

As a first step, we consider the energy change in the magnetic system due to
the interaction of a magnetic field  with the topological orbital magnetic 
moment that is induced by a non-coplanar chiral magnetic structure, characterized
by a nonzero scalar spin chirality $\hat{s}_i \cdot (\hat{s}_j \times
\hat{s}_k)$.
The free energy change (at $T = 0$ K) in the presence of an external field $\vec{B}$
is given by 
\begin{eqnarray}
\Delta {\cal E} &=& -\frac{1}{\pi} \mbox{Im}\,\mbox{Tr} \int^{E_F}
                    dE (E - E_F)\, G \hat{\cal H}_B G \;,
\label{Eq_Free_Energy-field-1}
\end{eqnarray}
with the perturbation operator to $\hat{\cal H}_B =
- \hat{\vec l} \cdot \vec{B}$, and $\hat{\vec l}$ the angular
momentum operator. This way we simplify the problem by accounting
  for the orbital moment associated with the electrons localized on 
the atoms and neglecting the 
contribution from the non-local component of the topological orbital
moment discussed, e.g., in Ref.\ [\onlinecite{HFN+16}]. 
  The reason why we are interesting here only in this part of the induced
  orbital moment is related to the interpretation of the TCI discussed
  previously \cite{MPE20}. In this case the TCI is characterized by the
  energy change due to the interaction of the spin magnetization with the
  orbital moment induced by the chiral magnetic configuration.

Next, we assume a non-collinear magnetic structure in the system, which
is treated as a perturbation leading to a change of the Green function
$G_0$ for collinear magnetic system according to:
\begin{eqnarray}
G &=& G_0 + G_0 V G_0 + G_0 V G_0 V G_0 + ... \;,
\label{Eq_Free_Energy-TCI-1}
\end{eqnarray}
%
where $V = V(q_1, q_2)$ is a perturbation due to the $2q$ spin
  modulation given by Eq.\ (\ref{spiral2})
  and  discussed in 
    Ref.\ [\onlinecite{MPE20}] when considering the three-spin exchange
    interaction parameters.
Using the expression in Eq.\ (\ref{Eq_Free_Energy-field-1}) we keep here
only the terms giving the three-site energy contribution corresponding to
the topological orbital susceptibility $\chi^{\rm{TO}}_{ijk}$:
\begin{eqnarray}
\Delta {\cal E}^{(3)} &=& -\frac{1}{\pi} \mbox{Im}\,\mbox{Tr} \int^{E_F}
                    dE (E - E_F)\,
   [ G_0 V G_0 V G_0 \hat{\cal H}_B G_0 \nonumber \\
  && + \,
                    G_0 V G_0 \hat{\cal H}_B G_0V G_0 + G_0 \hat{\cal H}_B G_0 V G_0 V G_0 ] \,.
\label{Eq_Free_Energy-TCI-2}
\end{eqnarray}
 As it is shown in the Appendix \ref{SEC:TOM}, Eq.\ (\ref{Eq_Free_Energy-TCI-2}) can
 be transformed to the form
\begin{eqnarray}
\Delta {\cal E}^{(3)} &=&  \frac{1}{3} \frac{1}{\pi} [ \mbox{Im}\,\mbox{Tr}
                    \int^{E_F} dE [ V G_0 V G_0\hat{\cal H}_B G_0
                    \nonumber \\
  &&+       V G_0 \hat{\cal H}_BG_0VG_0  + \hat{\cal H}_B G_0 V G_0 V G_0 ]    \;.
\label{Eq_Free_Energy-L-B}
\end{eqnarray}
This leads to an expression for the 3-spin topological orbital
susceptibility (TOS) responsible for the topological orbital moment (TOM)
induced in the trimer due to the magnetic configuration
characterized by a finite scalar spin chirality:
\begin{eqnarray}
  \chi^{\rm{TO}}_{ijk}  &=& - \frac {1}{4\pi} \mbox{Im}\, \mbox{Tr} \int^{E_F} dE \nonumber \\
 &&                    
\times \Big[ \underline{T}^{i, x}\, \underline{\tau}^{ij}
\underline{T}^{j, y}\, \underline{\tau}^{jk}
               \underline{l}_z^{k}\,  \underline{\tau}^{ki} 
- \underline{T}^{i, y}\, \underline{\tau}^{ij}
 \underline{T}^{j, x}\, \underline{\tau}^{jk}
     \underline{l}_z^{k}\,  \underline{\tau}^{ki}
               \nonumber \\       
&&     - \underline{T}^{i, x}\, \underline{\tau}^{ij}
\underline{l}_z^{j}\, \underline{\tau}^{jk}
               \underline{T}^{k, y}\, \underline{\tau}^{ki}\;
+ \underline{T}^{i, y}\, \underline{\tau}^{ij}
\underline{l}_z^{j}\, \underline{\tau}^{jk}
               \underline{T}^{k, x}\, \underline{\tau}^{ki} \;
               \nonumber \\       
&& +
    \underline{l}_z^{i} \, \underline{\tau}^{ij} 
\underline{T}^{j, x}\, \underline{\tau}^{jk}
               \underline{T}^{k, y}\, \underline{\tau}^{ki}\;
 - \underline{l}_z^{i}\,  \underline{\tau}^{ij}
\underline{T}^{j, y}\,\underline{\tau}^{jk}
               \underline{T}^{k, x}\, \underline{\tau}^{ki} \Big] \; .
               \nonumber \\       
\label{Eq:TOS} 
\end{eqnarray}
%
As it was mentioned above, the TOS given by Eq.\ (\ref{Eq:TOS}) characterizes the
  topological orbital moment along the $z$-axis in the global frame of
  reference, which is aligned with the magnetization direction of the  
 FM reference  system. 
 Moreover, for the system under consideration,
  with all magnetic atoms equivalent,
 one has  $L_1^{TO} = L_2^{TO} = L_3^{TO} = L^{TO}$ for the trimers
 $\Delta_1$  and  $\Delta_2$.   
This implies that the expression for the
TOS, $\chi^{\rm{TO}}_{ijk}$, given in Eq.\
(\ref{Eq:TOS}), gives access to the TOM  $L_i^{TO} = L_{i,z}^{TO}$ induced on 
atom $i$ that has its spin orientation $\hat{s}_i || \hat{z}$ for the FM
reference state.

\begin{figure}[h]
\includegraphics[width=0.35\textwidth,angle=0,clip]{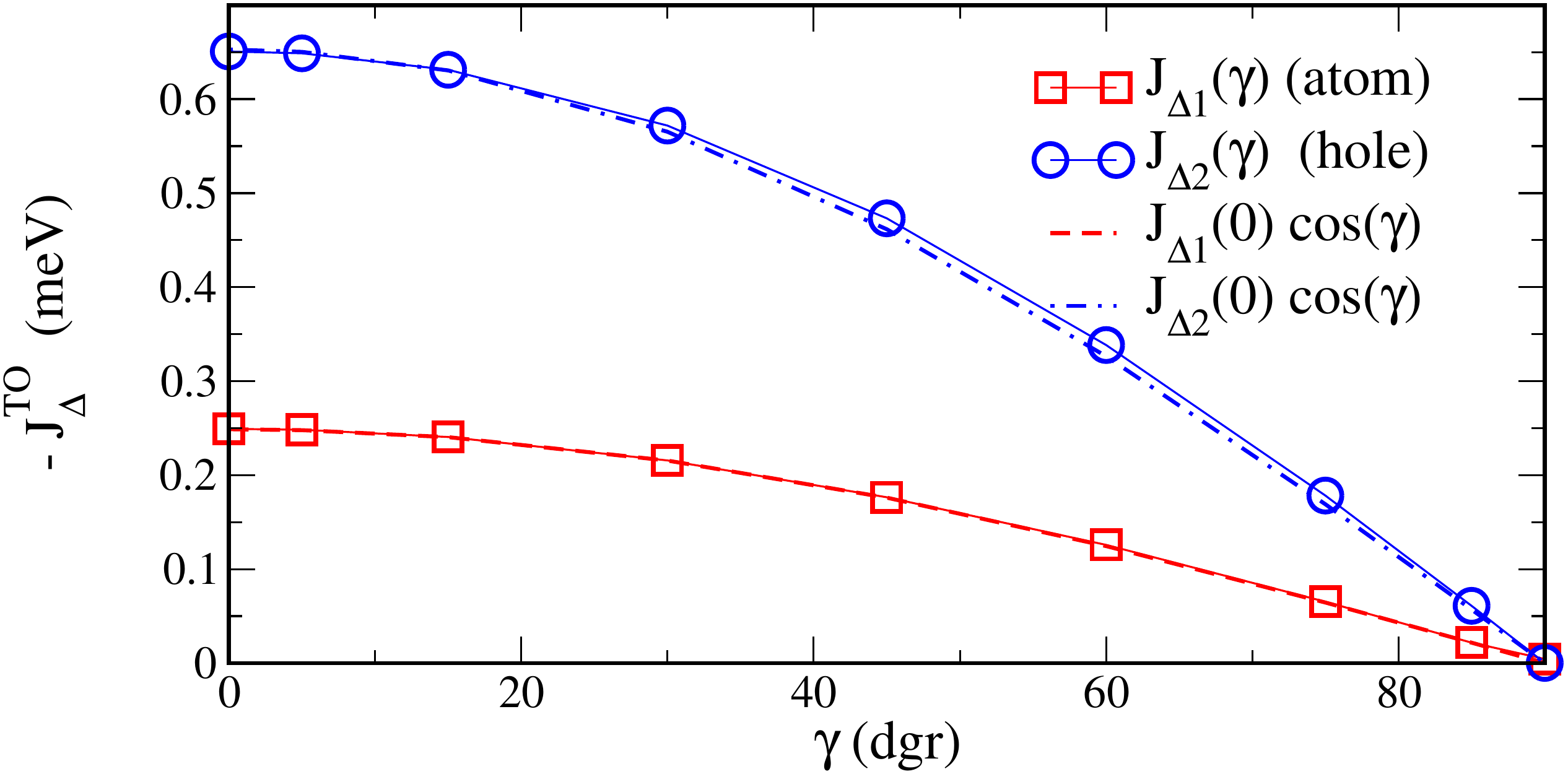}\;(a)\\
\includegraphics[width=0.35\textwidth,angle=0,clip]{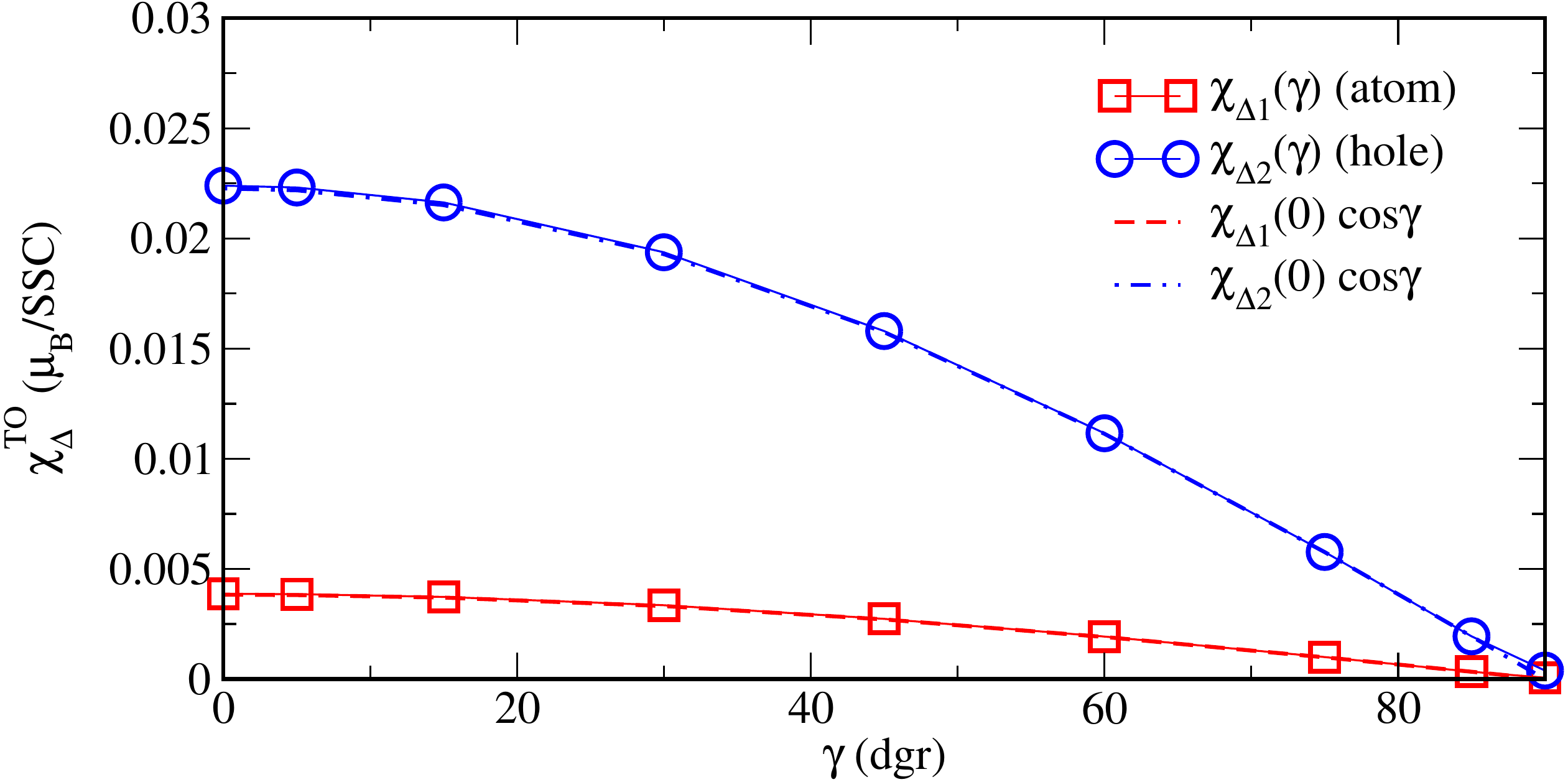}\;(b)
\includegraphics[width=0.35\textwidth,angle=0,clip]{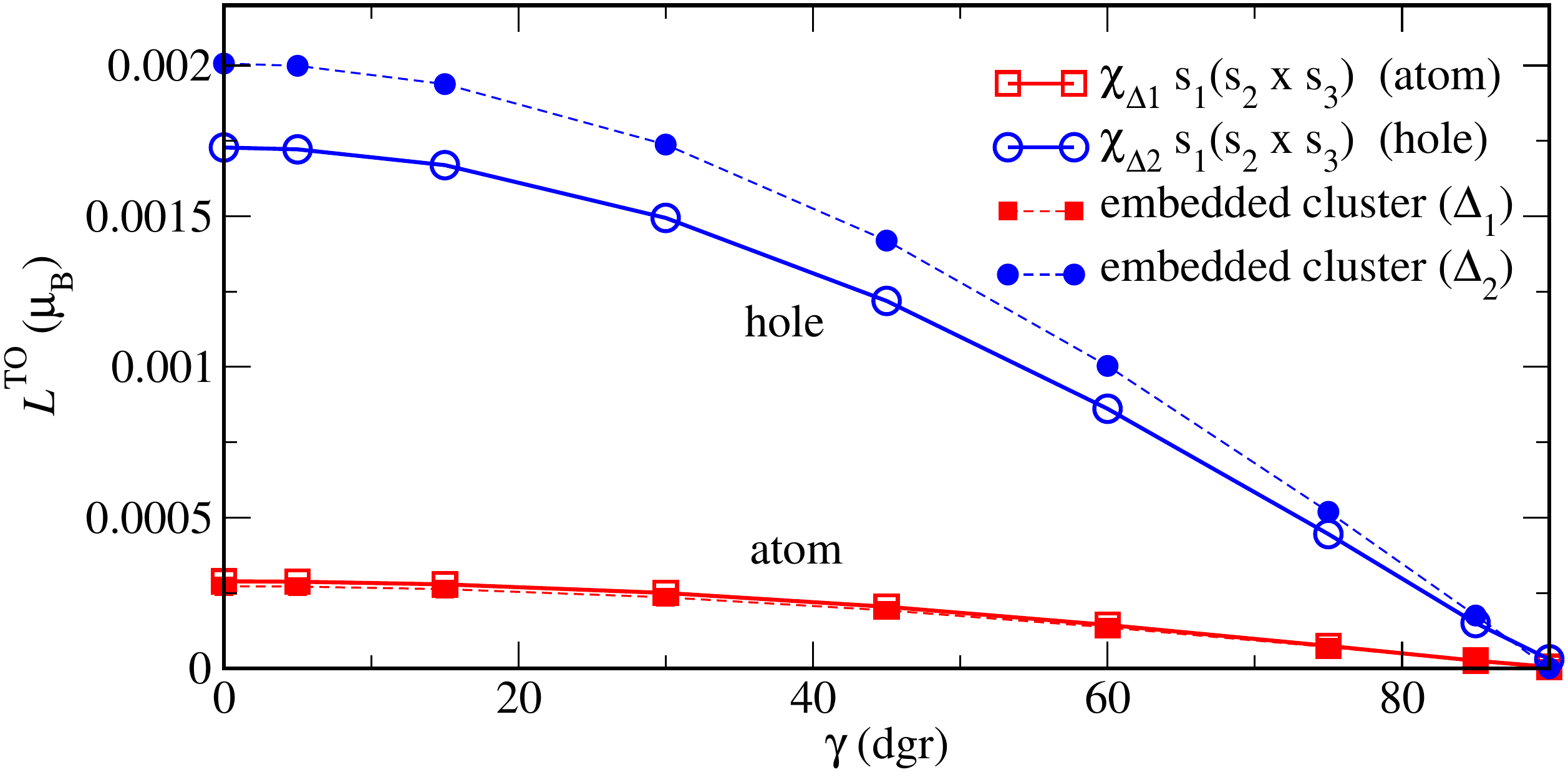}\;(c)
\includegraphics[width=0.35\textwidth,angle=0,clip]{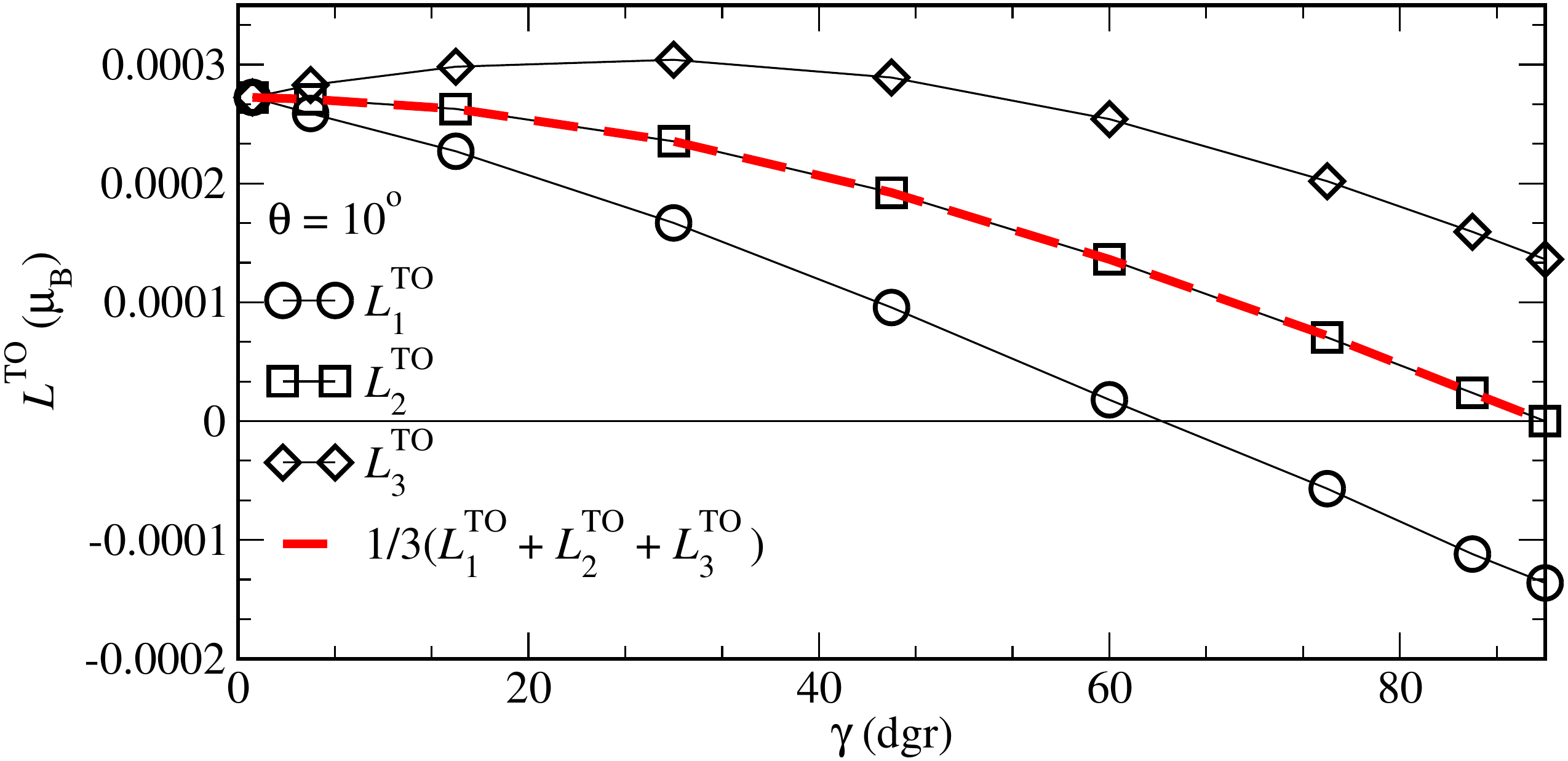}\;(d)
\caption{\label{fig:THEESPIN-Co-Ir} (a) Three-spin chiral exchange interaction
  parameters $J_{\Delta}(\gamma)$  and (b) topological orbital susceptibility (TOS, for SOC =
  0), calculated for Fe on Ir (111), as a function of the angle between
  the magnetization and normal $\hat{n}$ to the surface, for  the smallest
  triangles $\Delta_1$ and $\Delta_2$.  The dashed lines represent
  $J_{\Delta}(0) \, \cos( \gamma )$ (a) and
  $\chi^{\rm{TO}}_{\Delta}(0) \, \cos( \gamma )$ (b), respectively.
  To stress the relation between $J_{\Delta}$ and
  $\chi^{\rm{TO}}_{\Delta}$, we plot $-J_{\Delta}$ in panel (a).
  (c)  Topological orbital moment $L^{\rm{TO}}(\gamma)$
  (calculated for SOC = 0) induced by a 
  three-site chiral spin tilting by 
  $\theta = 10^\circ$, for trimers
  $\Delta_1$ (red squares, centered at an Ir atom) and 
  $\Delta_2$ (blue circles, centered by the hole in the Ir layer). The
  solid line represents the results obtained for $L^{\rm{TO}}(\gamma) =
  \chi^{\rm{TO}}_{\Delta} \hat{s}_i \cdot (\hat{s}_j 
  \times \hat{s}_k)$, while the dashed line represents the results
  of a direct calculations of $L^{\rm{TO}}(\gamma)$ for an embedded
  three-atomic Fe cluster.   (d) Orbital moments $L_i^{\rm{TO}}(\gamma)$
  on a three-atomic embedded Fe cluster $\Delta_1$ in Fe monolayer on Ir
  (111), induced at SOC = 0 due to the tilting of magnetic moments by 
  $\theta = 10^\circ$ with respect to the magnetization direction.
   The dashed line represents the average orbital moment.
    }  
\end{figure}

\begin{figure}
  \includegraphics[width=0.17\textwidth,angle=0,clip]{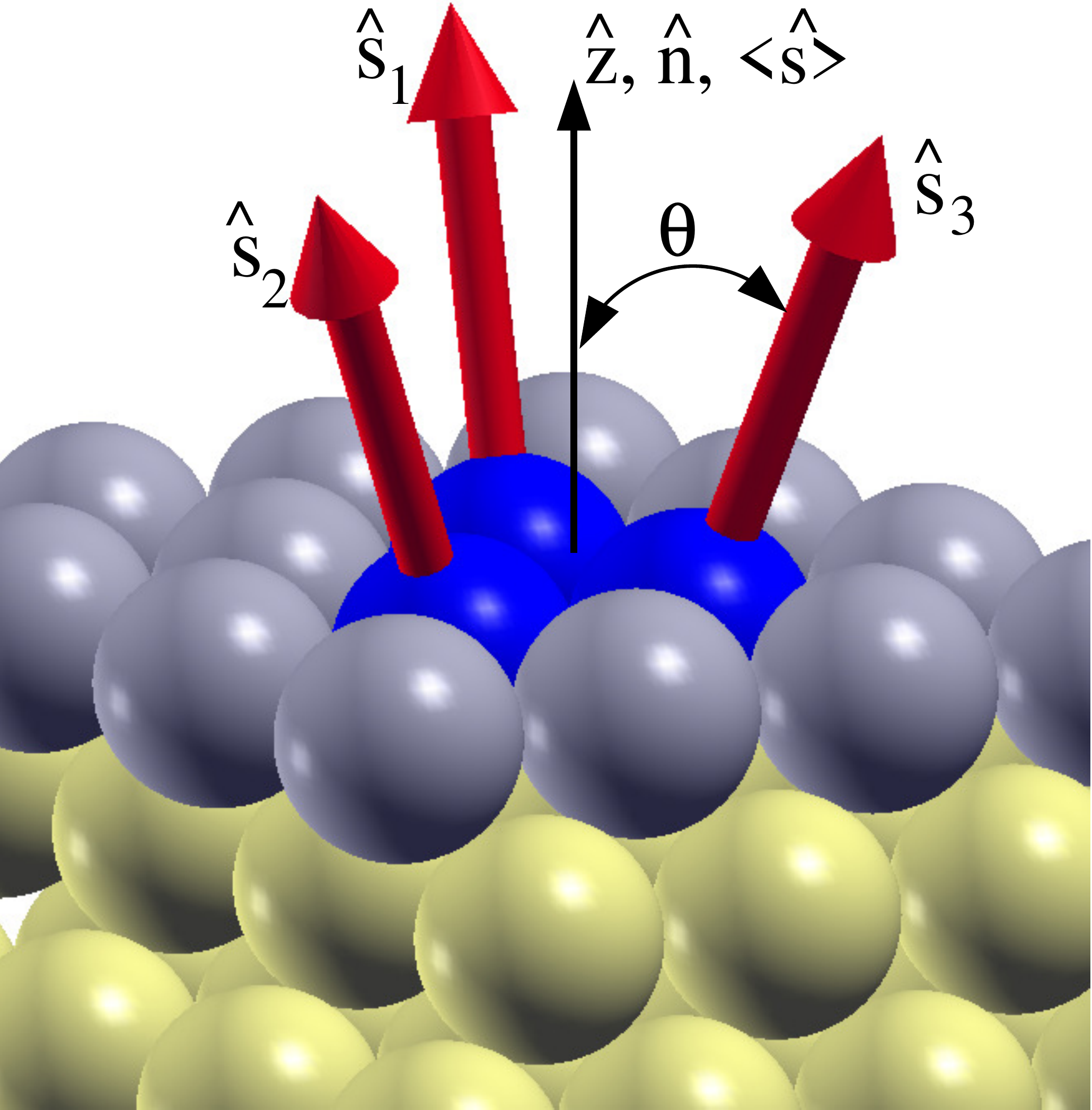} (a)\\
  \includegraphics[width=0.17\textwidth,angle=0,clip]{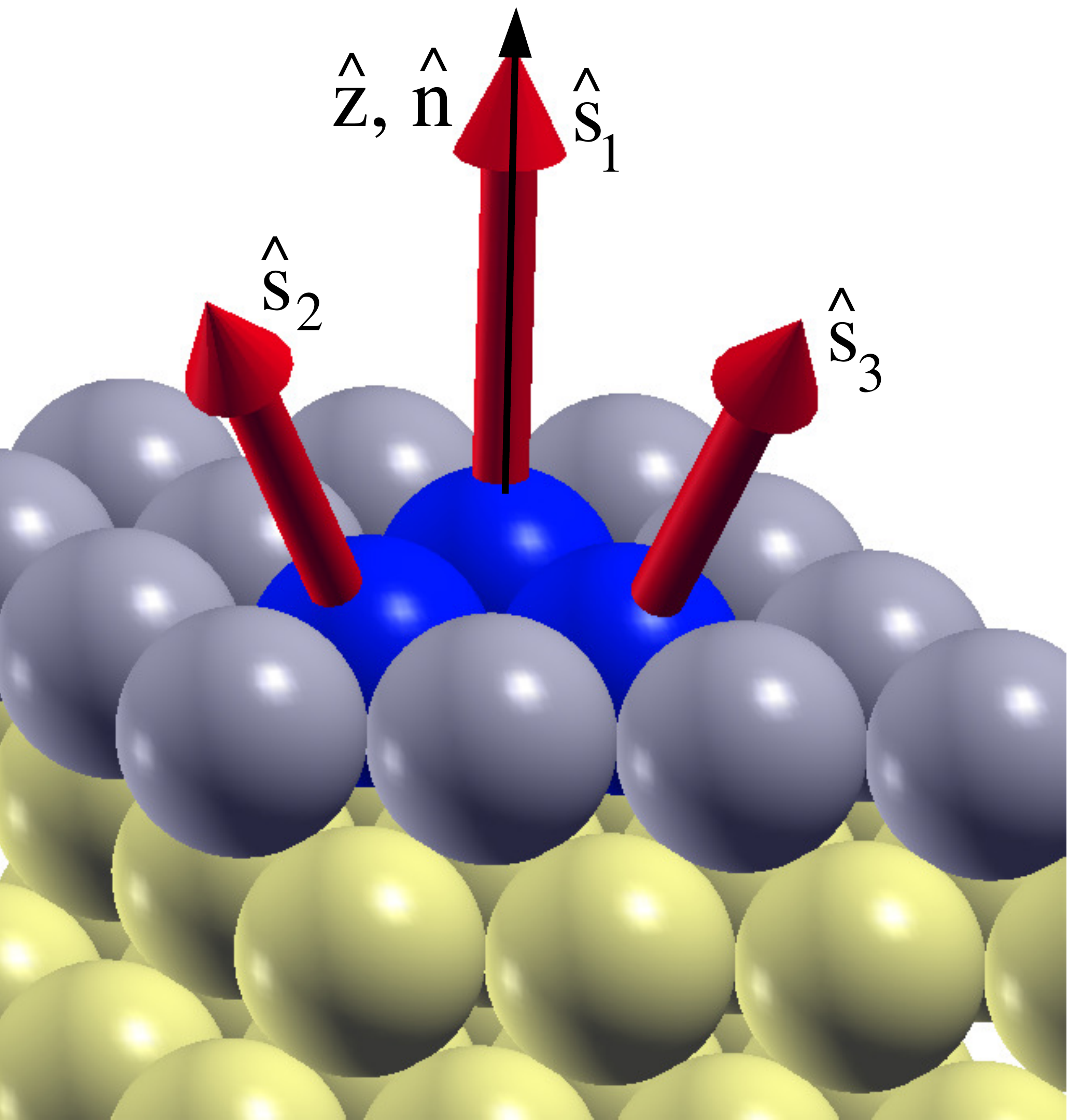} (b)
  \includegraphics[width=0.17\textwidth,angle=0,clip]{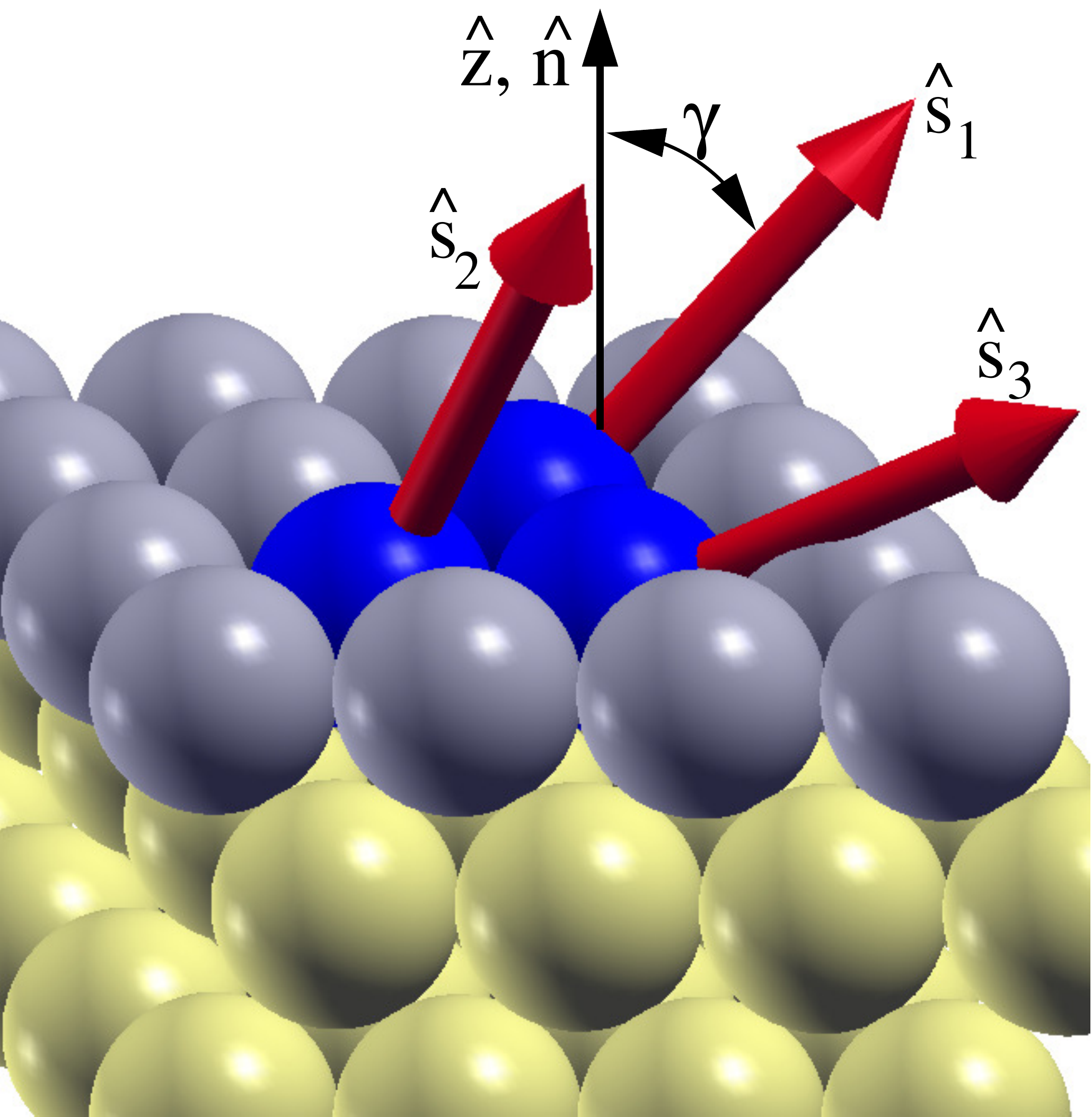} (c)
  \caption{\label{fig:Geometry} Spin orientation in three-atomic cluster
  in the magnetic monolayers on Ir (111): (a) $\theta$ - the angle of
  tilting tilting of magnetic moments $\hat{s}_i$ of the trimer with
  respect to the magnetization direction;
  (b) initial spin configuration used in the calculations of the TOM on
  the atom $i$, for the embedded cluster; (c) $\gamma$ - the angle between
  the direction of spin magnetic moment  
  $\hat{s}_i$ and the normal to the surface $\hat{n}_i$,
 within the plane perpendicular to the rotation axis. 
  }  
\end{figure}

Using the expression
in Eq.\ (\ref{Eq:TOS}), calculations of the three-spin topological
orbital susceptibility together with the TCI has been performed for 1ML
of $3d$ metals on Ir (111) surface. Corresponding values calculated for
a Fe overlayer are 
represented in Fig.\ \ref{fig:THEESPIN-Co-Ir} as a function of the 
angle $\gamma$ between the magnetization direction and the normal $\hat{n}$ to the
surface plane (see Fig.\  \ref{fig:Geometry}).
%
%
%
%
These results give evidence for
the common dependence of the TCI and the TOS on the flux of the spin
magnetization through the triangle area. 
The calculations have been performed for the two smallest trimers,
$\Delta_1$ and $\Delta_2$, centered at the Ir atom and the hole site
in the  Ir surface layer, respectively (Fig.\  \ref{fig:geometry_triangles}). As
both quantities, $J_{ijk}$ and  $\chi^{\rm{TO}}_{ikj}$ , follow the permutation properties of 
the product $\hat{s}_i \cdot (\hat{s}_j \times \hat{s}_k)$, we introduce
the quantities $J_{\Delta} = J_{ijk} -  J_{ikj}$ and  $\chi^{\rm{TO}}_{\Delta} =
\chi^{\rm{TO}}_{ijk} -  \chi^{\rm{TO}}_{ikj}$, which allow to
avoid double summation over the counter-clockwise and counter-anticlockwise 
contributions upon a summation over the lattice sites in the energy or
orbital moment calculations.
 Note also that in the present case with all magnetic atoms
  equivalent  $J_{\Delta} = J_{ijk} -  J_{kji} = 
  J_{ijk} -  J_{jik}$ as well as  $\chi^{\rm{TO}}_{\Delta} = \chi^{\rm{TO}}_{ijk}
  -  \chi^{\rm{TO}}_{kji} = \chi^{\rm{TO}}_{ijk} -  \chi^{\rm{TO}}_{jik}$. 
As one can see, both $J_{\Delta}(\gamma)$ and $\chi^{\rm{TO}}_{\Delta}(\gamma)$
are in  perfect agreement with the functions $J_{\Delta}(0)
\cos( \gamma )$ and  $\chi^{\rm{TO}}_{\Delta}(0) \cos( \gamma )$ in line with 
Eq.\ (\ref{Eq_INTERPR-4}) to be considered here 
for the situation $\hat{m} \| \hat{z}$, i.e.\
$\hat{m} \cdot \hat{\vec{l}} \sim \cos( \gamma )$.

The  result for  $\chi^{\rm{TO}}_{\Delta}$ can be compared  to the pure
 topological orbital moment (TOM) $L^{TO}(\gamma)$ 
 that is derived directly from the electronic structure 
 when the SOC is suppressed. Corresponding 
 calculations have been done for 
 3-atom Fe clusters 
($\Delta_1$ or $\Delta_2$, as is shown in Fig.\ \ref{fig:geometry_triangles})
embedded in a Fe monolayer on the Ir(111) surface.
For this, the   Fe spin
moments $\hat{s}_1, \hat{s}_2$, and $ \hat{s}_3$ 
of the cluster 
given by 
$\hat{s}_i = (\sin(\theta)\, \cos(\phi_i), \sin(\theta) \,\sin(\phi_i), \cos(\theta))$
(with $\phi_{i+1} -\phi_i = 120^\circ$)
are 
 tilted by the angle $\theta$ with
respect to the 'average' spin direction $\langle \hat{s} \rangle = 1/3(\hat{s}_1 +
\hat{s}_2 + \hat{s}_3)$, as it is shown in 
 Fig.\ \ref{fig:Geometry}.

It should be emphasized 
 that a one-to-one comparison of two approaches is
  only sensible when performing the
  corresponding calculations under identical conditions.
  This implies here an orientation of the spin magnetic moment
  $\hat{s}_i$ as well as of
   the topological orbital moment $\hat{L}^{\rm{TO}}_i =
  \hat{L}^{\rm{TO}}_{i,z}$ on the atom $i$ along the global $\hat{z}$
  axis, i.e. $\hat{L}^{\rm{TO}}_i || \hat{s}_i || \hat{z}$,
  identical to the conditions used within 
  the perturbational approach.
  In the calculations for the embedded cluster with  finite spin tilting angles, 
  this condition can be met only for one atom of the
  trimer at a time, e.g.\ for  atom $1$ 
  (see Fig.\ \ref{fig:Geometry} (b)).  
    The angle $\gamma$ characterizes the relative orientation of the spin
  direction $\hat{s}_1$ and the normal $\hat{n}$ to the surface (i.e.\ the plane of the
  triangle), as is shown in Fig.\ \ref{fig:Geometry} (c).
  The
  initial spin configuration ($\gamma = 0$) used in the embedded cluster  
  calculations shown in Fig.\  \ref{fig:Geometry} (b) can be obtained from the
  configuration shown in Fig.\ \ref{fig:Geometry} (a)
  by a  corresponding rotation ${\cal R}_1$ 
  according to
  $\hat{s}_1^{(b)} = {\cal R}_1\hat{s}_1^{(a)}$,
  $\hat{s}_2^{(b)} = {\cal R}_1\hat{s}_2^{(a)}$, and
  $\hat{s}_3^{(b)} = {\cal R}_1\hat{s}_3^{(a)}$, such that
  ${\cal R}_1\hat{s}_1^{(a)} || \hat{z}$. 
  The TOMs of atoms 2 and 3,   
   $L_2^{\rm{TO}}(\gamma)$ and 
  $L_3^{\rm{TO}}(\gamma)$, respectively, and their
   dependency on the angle $\gamma$ can be
obtained from corresponding spin configurations obtained 
by applying  the rotations 
${\cal  R}_2$ and  ${\cal R}_3$, 
fixed by the requirement ${\cal  R}_2\hat{s}_2^{(a)} || \hat{z}$ 
and ${\cal R}_3\hat{s}_3^{(a)} || \hat{z}$, respectively.

 As one  notices 
from  Fig.\ \ref{fig:THEESPIN-Co-Ir} (d)
for the cluster $\Delta_1$, 
 the variation of the calculated
  TOM $L_i^{\rm{TO}}(\gamma)$
with the tilting angle  $\gamma$ is somewhat  different 
for the various atoms in the cluster, 
with the difference increasing with increasing  $\theta$.
 Fig.\ \ref{fig:THEESPIN-Co-Ir} (c) represents 
 by full symbols the dependence
of the corresponding averaged 
TOM $L^{\rm{TO}}(\gamma) =  1/3(L^{\rm{TO}}_1 + L^{\rm{TO}}_2 +
L^{\rm{TO}}_3)$  on the angle $\gamma$,
with $L^{\rm{TO}}_i(\gamma)$ induced due to the tilting of all spin
moments by the angle $\theta = 10^o$ (see Fig.\  \ref{fig:Geometry}).
Rotating the lattice, the direction of the normal vector $\hat{n}$ rotates
with respect to the fixed $\hat{z}$ axis, leading to an
increase of the angle $\gamma$  and to a decrease of the topological
orbital moment of the 3-atomic cluster.

As one can see in Fig.\ \ref{fig:THEESPIN-Co-Ir} (c), 
for both considered clusters
the results for $L^{\rm{TO}}(\gamma)$   are 
in good agreement with the TOM (given by open symbols) evaluated via
$L_{\Delta}^{\rm{TO}}(\gamma) = \chi^{\rm{TO}}_{\Delta} 
\hat{s}_i \cdot (\hat{s}_j \times \hat{s}_k)$ 
on the basis  the TOS plotted in
Fig.\ \ref{fig:THEESPIN-Co-Ir} (b). 
These findings clearly support the concept of the
topological orbital susceptibility as well
as the interpretation of the  
 topological orbital moment. 

\medskip

\begin{figure}
  \includegraphics[width=0.43\textwidth,angle=0,clip]{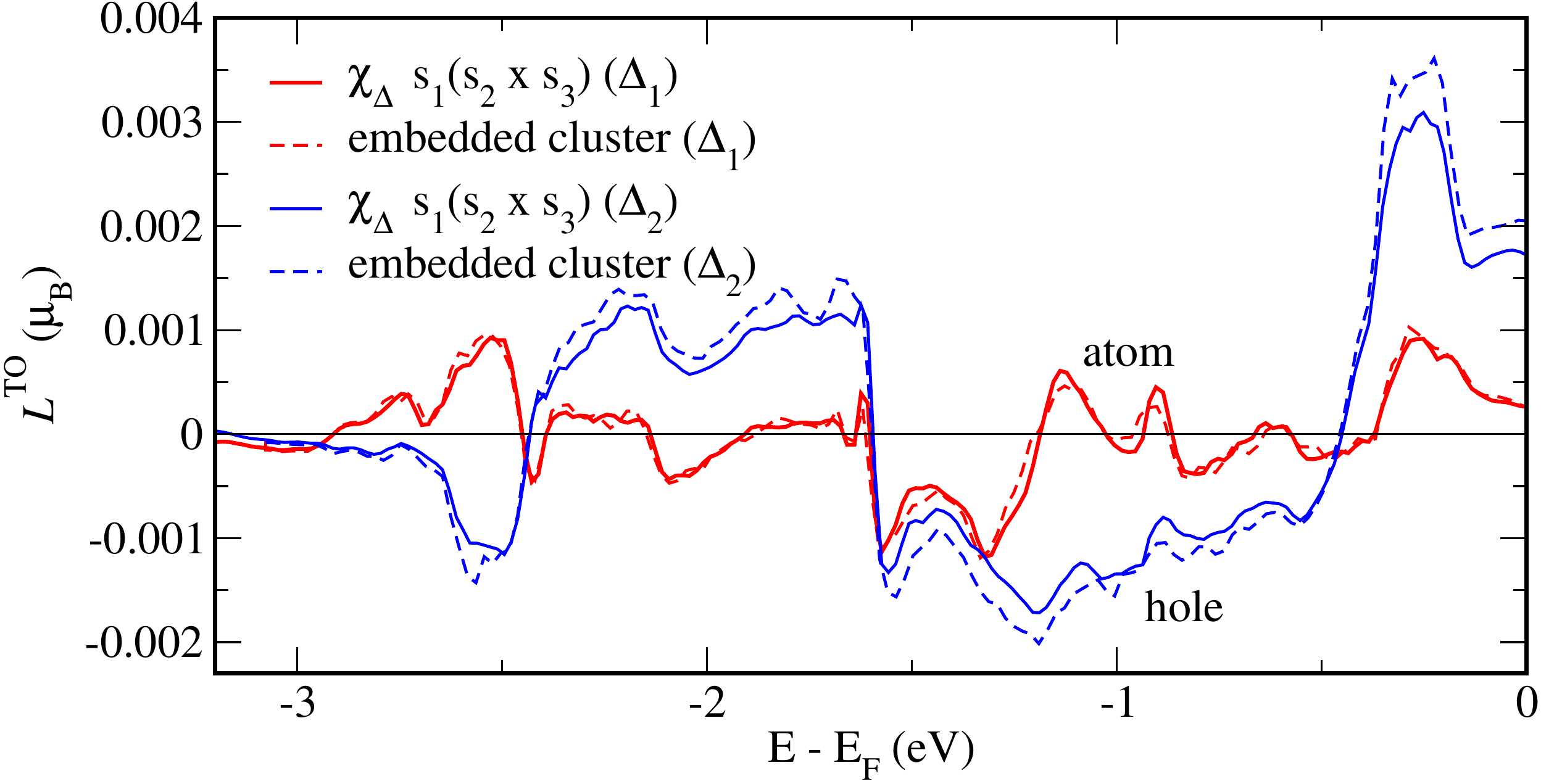} \,(a)\\ 
  \includegraphics[width=0.43\textwidth,angle=0,clip]{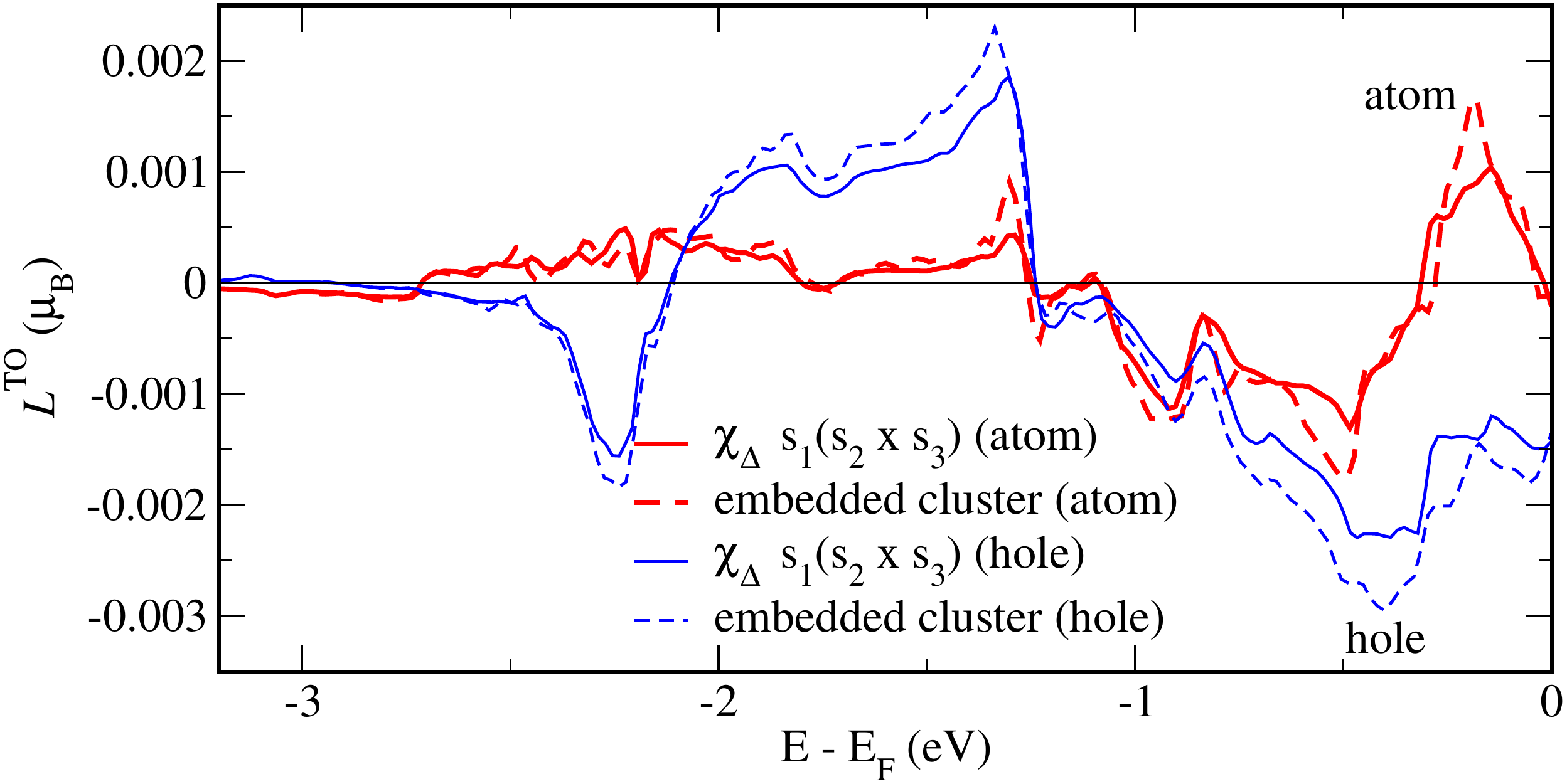} \,(b)
\caption{\label{fig:SSC-OM_EF}  Topological orbital moment (calculated
  for SOC = 0) induced by a three-site chiral spin tilting by $\theta =
  10^\circ$, for the smallest triangles $\Delta_1$ (red, centered at an Ir atom) and
  $\Delta_2$ (blue, centered by the hole in the Ir layer) in 1 ML of Fe (a) and Mn
  (b) on Ir (111), as a function of the occupation. The solid line represents
  the results obtained for $\chi^{\rm{TO}}_{\Delta} \hat{s}_i \cdot (\hat{s}_j
  \times \hat{s}_k)$, while the dashed line represents the results
  directly calculated for the embedded three-atomic Fe cluster.  
  }  
\end{figure}

 In Fig.\ \ref{fig:SSC-OM_EF} we represent 
 in addition the dependence of the
 directly calculated TOM on the occupation of the electronic states,
 which is considered for 
 embedded Fe and Mn three-atomic clusters $\Delta_1$ and $\Delta_2$,
 for 1ML Fe (a) and 1ML Mn (b), respectively, on the Ir (111)
 surface.
 Such an energy-resolved representation may allow to monitor differences
 for the various quantities considered concerning their origin in the
 electronic structure.  
 The TOM plotted in Fig.\ \ref{fig:SSC-OM_EF} by a dashed line is
 calculated for atom 1 in the embedded cluster with magnetic moments
 on the atoms tilted by $\theta = 10^\circ$ with respect to the 
 average spin direction $\langle \hat{s} \rangle$. In turn, $\langle
 \hat{s} \rangle$ is tilted by $\gamma = 10^\circ$ to have an orientation of
 $\hat{s}_1$ along the normal $\hat{n}$ to the surface (see Fig.\
 \ref{fig:Geometry} (b)). 
The solid line represents the orbital moment calculated using the three-spin
 topological orbital susceptibility $\chi^{\rm{TO}}_{\Delta}$ 
 scaled by the scalar spin chirality factor, i.e.\ $L^{\rm{TO}}_{\Delta} =
 \chi^{\rm{TO}}_{\Delta} \hat{s}_1 \cdot (\hat{s}_2 \times \hat{s}_3)$. 
As one can see, both results are close to each other and the difference
can be attributed to a finite $\theta$ angle in the case of the embedded
cluster calculations.
Comparing the results for Fe and Mn, one can also see that the different
sign of the induced topological orbital moments on Fe and Mn atoms
(this implies $E - E_F = 0$ in Fig.\ \ref{fig:SSC-OM_EF}) is mainly a
result of a different occupation of the electronic states in these materials. 
%
%
Thus, the presented results give clear evidence that the susceptibility
calculated using Eq.\ (\ref{Eq:TOS}) characterizes an orbital moment on
the atoms as a response to the effective magnetic field represented by
scalar spin chirality for every three-atomic cluster.
It has a form closely connected to that for the TCI and
therefore should be seen as a source for the TCI according 
to Eq.\ (\ref{Eq_INTERPR-2}).

\subsection{Topological spin susceptibility  (TSS)
 \label{TSS}}

\begin{figure}
\includegraphics[width=0.43\textwidth,angle=0,clip]{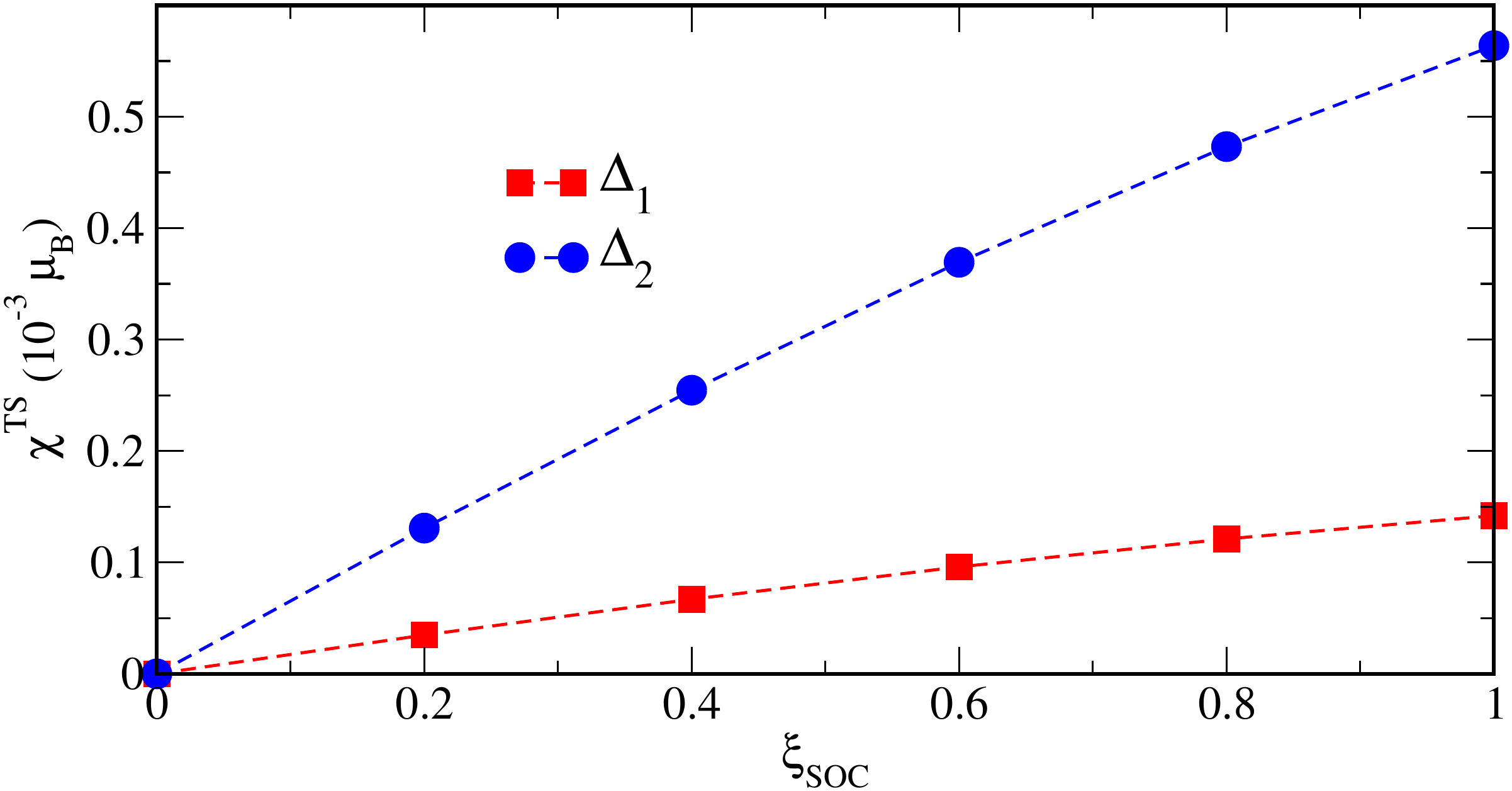}\;(a)
\includegraphics[width=0.43\textwidth,angle=0,clip]{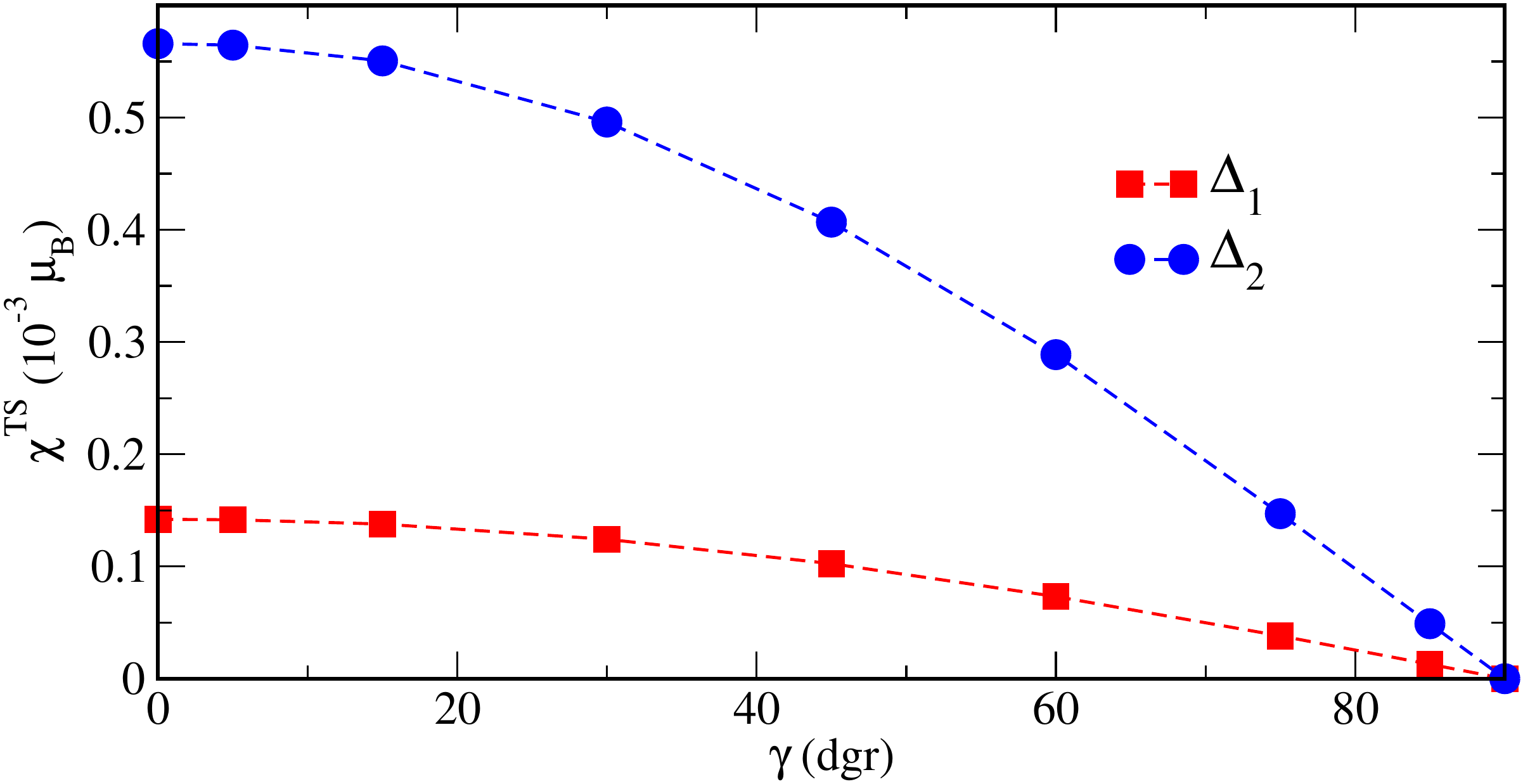}\;(b)
\caption{\label{fig:TOPO-SPIN-GAMMA}'Topological' spin susceptibility
  $\chi^{\rm{TS}}_{\Delta}$ 
  calculated for trimers $\Delta_1$ and $\Delta_2$ in Fe on Au (111):
  (a) as a function of $\xi_{\rm{SOC}}$ and (b) as a function of the
  angle between  the magnetization and normal $\hat{n}$ to the surface.
  }  
\end{figure}

  Discussing in Section \ref{TCI+SOC} the role of the SOC for the TCI
   using a description of the electronic structure in terms of
  the non-relativistic Green functions, we have  shown its
  responsibility for the induced spin magnetic moment in the presence of
  a chiral magnetic structure governing the TOM. The appearance of the
  spin moment induced due to a chiral magnetic structure can also be
  demonstrated explicitly within the fully relativistic approach,
  that gives access to an estimate for the TCI based on
  Eq.\ (\ref{Eq_INTERPR-4}).
  This can be done by introducing a
   'topological' spin susceptibility (TSS) 
   in analogy to the 
   TOS that is given explicitly
    by the expression in Eq.\ (\ref{Eq:TOS}).
 Following the discussion of the role of the  SOC for the
  induced spin magnetization
   given by Eq.\ (\ref{Eq_T-spin})
   that is based on a 
  non-relativistic reference system
one obviously has to account for the SOC 
when dealing with the   TSS 
$\chi^{\rm{TS}}_{ijk}$.
This is done here by working 
  on a  fully relativistic level
and representing the underlying
electronic structure in terms of the
retarded Green function evaluated by means
of the multiple-scattering formalism (see above). 
 This approach allows to write
  for  $\chi^{\rm{TS}}_{ijk}$ the expression:  
%
\begin{eqnarray}
  \chi^{\rm{TS}}_{ijk}  &=& - \frac {1}{4\pi} \mbox{Im}\, \mbox{Tr} \int^{E_F} dE \nonumber \\
 &&                    
\times \Big[ \underline{T}^{i, x}\, \underline{\tau}^{ij}
\underline{T}^{j, y}\, \underline{\tau}^{jk}
               \underline{\sigma}_z^{k}\,  \underline{\tau}^{ki} 
- \underline{T}^{i, y}\, \underline{\tau}^{ij}
 \underline{T}^{j, x}\, \underline{\tau}^{jk}
     \underline{\sigma}_z^{k}\,  \underline{\tau}^{ki}
               \nonumber \\       
&&     - \underline{T}^{i, x}\, \underline{\tau}^{ij}
\underline{\sigma}_z^{j}\, \underline{\tau}^{jk}
               \underline{T}^{k, y}\, \underline{\tau}^{ki}\;
+ \underline{T}^{i, y}\, \underline{\tau}^{ij}
\underline{\sigma}_z^{j}\, \underline{\tau}^{jk}
               \underline{T}^{k, x}\, \underline{\tau}^{ki} \;
               \nonumber \\       
&& +
    \underline{\sigma}_z^{i} \, \underline{\tau}^{ij} 
\underline{T}^{j, x}\, \underline{\tau}^{jk}
               \underline{T}^{k, y}\, \underline{\tau}^{ki}\;
 - \underline{\sigma}_z^{i}\,  \underline{\tau}^{ij}
\underline{T}^{j, y}\,\underline{\tau}^{jk}
               \underline{T}^{k, x}\, \underline{\tau}^{ki} \Big] \; .
               \nonumber \\       
\label{Eq:TSS} 
\end{eqnarray}
%
Using this expression,
$\chi^{\rm{TS}}_{\Delta} = \chi^{\rm{TS}}_{ijk} -                    
                           \chi^{\rm{TS}}_{ikj}$ 
  was calculated as a function of 
the SOC scaling parameter
 $\xi_{\rm{SOC}}$, as well as 
  the angle $\gamma$ defined above,
   for 1ML Fe on Au(111) surface.
Fig.\  \ref{fig:TOPO-SPIN-GAMMA} (a) represents the dependence of
 $\chi^{\rm{TS}}_{\Delta}$ on  $\xi_{\rm{SOC}}$,
 clearly  demonstrating the relativistic origin of
this quantity giving rise to the corresponding contribution to the TCI
(see Eq.\   
(\ref{Eq_T-spin})). These results can be compared with the
$\xi_{\rm{SOC}}$-dependence of the TCI plotted in Fig.\  \ref{Eq:J_XYZ}.
 As it has been seen in Fig.\ \ref{fig:THEESPIN-Co-Ir},  
when comparing
$J_{\Delta}$ and $\chi^{\rm{TO}}_{\Delta}$, 
one finds 
a different sign for
$J_{\Delta}$ and $\chi^{\rm{TS}}_{\Delta}$. 

On the other hand, $\chi^{\rm{TS}}_{\Delta}$ should follow 
the angle $\gamma$ between the magnetization 
and surface normal $\hat{n}$, as it has been obtained for $\chi^{\rm{TO}}_{\Delta}$. As one can
see in Fig.\  \ref{fig:TOPO-SPIN-GAMMA} (b), $\chi^{\rm{TS}}_{\Delta}(\gamma)$
is well represented by $\chi^{\rm{TS}}_{\Delta}(0) \, \cos(\gamma)$,
demonstrating a common behavior of the topological spin susceptibility
$\chi^{\rm{TS}}_{\Delta}$ and the TOS $\chi^{\rm{TO}}_{\Delta}$.

Using the results in Fig.\  \ref{fig:TOPO-SPIN-GAMMA} (a) together with the 
 ground state spin moment $m_{\rm{Fe}} =
3 \mu_{\rm B}$ and the approximate exchange splitting $\Delta_{\rm{xc}} \sim 3$ eV
calculated for 1 ML Fe /Au(111), one can give 
the  crude 
estimate  $1 {\rm eV}/\mu_{\rm B}$  for the
effective $B$-field $B_{\rm{eff}} \approx \Delta_{\rm{xc}}/m_{\rm{Fe}}$
giving access to the TCI
 connected with the TSS $\chi^{\rm{TS}}_{\Delta}$. 
Using Eq.\ (\ref{Eq_INTERPR-2})
approximated by $J_{\Delta} \approx 
B_{\rm{eff}} \, \chi^{\rm{TS}}_{\Delta}$, one obtains
the values 
 $J_{\Delta_1}  \approx
-0.14$  and $J_{\Delta_2}  \approx -0.57$ meV for $\xi_{\rm{SOC}} = 1$,
which are in reasonable agreement with the 
properly calculated values 
shown in Fig.\ \ref{fig:THEESPIN-SOC}
supporting the concept of a TSS as introduced here.
%
%

\section{Summary}

  To summarize, we have stressed that the TCI derived in Ref.\ [\onlinecite{MPE20}] is fully in line with the symmetry properties of a fully
  antisymmetric rank-3 tensor, specific only for this type of
  interaction. This interaction should be distinguished
  from the 4-spin DMI-like exchange interactions obtained in different
  order of perturbation theory and characterized by different properties
  with respect to a permutation of the spin indices.

 We suggest an interpretation of the TCI showing its dependence on the
  relativistic SOC and on the TOS as a 
  possible source. 
Concerning the SOC, an analytical expression
based on a perturbative treatment of the SOC
as well as numerical results for the TCI parameter
demonstrate the role of the SOC as an ultimate source
for a non-zero TCI.   
 An expression 
 for the TOS 
that  reflects the topological origin of the TOS
and that is very similar to that for the TCI
parameters
  has been derived. 
Numerical results again demonstrate the intimate
connection between both quantities.

To allow for a more detailed discussion
of the TCI, the 
'topological' spin susceptibility (TSS)
has been  introduced as a quantity
 that reflects the impact of the SOC
   in the presence of a non-collinear 
   magnetic structure,
   leading to a non-vanishing TCI.
Corresponding numerical results also 
  demonstrated for the TSS its
  connection with the TCI parameters.

In summary,
the work presented not only 
revealed details of the mechanism giving rise
to the TCI and its connection  with related 
quantities but also 
clearly rebutted the misleading
  criticism raised by dos Santos Dias et al.\ \cite{SBL+21a}.

 \section{Acknowledgement}
Financial support by the DFG via SFB 1277 (Emergent Relativistic Effects
in Condensed Matter - From Fundamental Aspects to Electronic
Functionality) is gratefully acknowledged.

\appendix

\section{Properties of the TCI
  with respect to time reversal} \label{SEC:TR}

In the following  we address
the properties  of the TCI with respect to time reversal.
Considering the  TCI parameter
as a 'conventional' scalar quantity invariant with respect
to time reversal together with the
 scalar spin chirality 
 $\hat{s}_i\cdot (\hat{s}_j \times  \hat{s}_k) $ in the
 third term of the spin Hamiltonian
 given in Eq.\
(\ref{Eq_Heisenberg_general}) being antisymmetric with respect
to this operation  one is
erroneously  led  to the conclusion that the TCI
should not contribute to the energy expansion
in Eq.\ (\ref{Eq_Heisenberg_general}).

Concerning this,
we remind here that the expression
for the corresponding TCI parameter
was derived in Ref.\ [\onlinecite{MPE20}]
by considering spin tiltings as a perturbation to
a FM reference system that lead to a non-collinear spin modulation.
Using a representation of the electronic structure
of the FM state
by means of its Green function  $G_0$
the change in energy in second-order
with respect to the  perturbation is given by \cite{MPE20}
%
\begin{eqnarray}
\Delta {\cal E} &=& -\frac{1}{\pi} \mbox{Im}\,\mbox{Tr} \int^{E_F}
                    dE (E - E_F)\, G_0\, \Delta V\, G_0\, \Delta V\, G_0 \;, \nonumber \\
\label{Eq_Free_Energy-TCI-1A}
\end{eqnarray}
with the perturbation $\Delta V$ %
\begin{eqnarray}
 \Delta V &=&  \sum_i \beta \big( \vec{\sigma}\cdot\hat{s}_i
  -  \sigma_z\big) B_{xc}\;.
\label{Eq_perturbA}
\end{eqnarray}
%
and all spatial arguments and corresponding integrals omitted.

As the change in energy due to an arbitrary perturbation
should be invariant w.r.t.\ time reversal, this should
hold in particular for the term 
given in  Eq.\ (\ref{Eq_Free_Energy-TCI-1A})
that is second order concerning the  perturbation $\Delta V$.
This property is demonstrated in the following
making use of the
simplified version of Eq.\
(\ref{Eq_Free_Energy-TCI-1A}) 
\begin{eqnarray}
\Delta {\cal E} &=& -\frac{1}{\pi} \mbox{Im}\,\mbox{Tr} \int^{E_F}
                    dE \, 
                    (\vec{B}_{xc} \cdot \vec{\sigma})  \nonumber \\ 
   && \times G_0\,\xi_{SOC}(\vec{\sigma} \cdot
                    \hat{\vec l}) \, G_0\, \Delta V\, G_0\, \Delta V\, G_0
                    \;,
\label{Eq_Free_Energy-TCI-2A}
\end{eqnarray}
as given in  Eq.\ (\ref{Eq_INTERPR-4}),
with SOC  treated as a perturbation.
Treating the FM state otherwise on a non-relativistic level
its Green function $ G_0$ is block-diagonal w.r.t.\ the spin index:
\begin{eqnarray}
  G_0 &=& \frac{1}{2}(G_0^{\uparrow \uparrow} + G_0^{\downarrow \downarrow}) +
        \frac{1}{2}(G_0^{\uparrow \uparrow} - G_0^{\downarrow
        \downarrow})\sigma_z \nonumber \\
   &=& G_{01} \sigma_0 + G_{02} \sigma_z \;,
\label{Eq_Append_NRGF}
\end{eqnarray}
with $\sigma_0$ the unit $2 \times 2$ matrix  and the perturbation
\begin{eqnarray}
 \Delta V &=&  \sum_i ({\sigma}_x s_{x}^i + {\sigma}_y
                     s_{y}^i + {\sigma}_z (s_{z}^i - 1)) B_{xc}    \;.
\label{Eq_perturb2A}
\end{eqnarray}
Inserting Eqs.\ (\ref{Eq_Append_NRGF}) and  (\ref{Eq_perturb2A})
into
Eq.\  (\ref{Eq_Free_Energy-TCI-2A})
leads,  among others, to contributions of the form
%
\begin{eqnarray}
&\sim&  \mbox{Im}\,\mbox{Tr} [ (B_{xc}^i \sigma_z) (G_{01} \sigma_0)\, \xi_{SOC}^i(\sigma_z 
\hat{l}_z) \nonumber \\
 && \times  G_{01} \sigma_0 (\sigma_x B_{xc}^j) \, (G_{01} \sigma_0) 
      (\sigma_y B_{xc}^k) (G_{02} \sigma_z)]  \, , \nonumber
\end{eqnarray}
which are invariant w.r.t.\ time reversal,
as it will be shown next.
A corresponding application of
the time reversal operator ${\cal T} = -i \sigma_y {\cal K}$,
  with $K$ the operator of complex conjugation,
  to the Green function leads to
\begin{eqnarray}
  {\cal T}  (G_{01} \sigma_0 + G_{02} \sigma_z)  {\cal T}^{-1} =
  (G_{01}^* \sigma_0 - G_{02}^* \sigma_z)\,, 
\label{Eq_TR_GF}
\end{eqnarray}
while the operators
$ \Delta V$ and $(\sigma_z \hat{l}_z)$ are invariant under this
operation.
Accordingly, one has for the contribution given above:
%
\begin{eqnarray}
 &\sim& \mbox{Im}\,\mbox{Tr} [  (B_{xc}^i \sigma_z) (G^*_{01} \sigma_0)\, \xi_{SOC}^i (\sigma_z 
      \hat{l}_z) \nonumber \\
 && \times   G^*_{01} \sigma_0 (\sigma_x B_{xc}^j) \, (G^*_{01}
  \sigma_0) (\sigma_y B_{xc}^k) (-G^*_{02} \sigma_z) ]\,. \nonumber
\end{eqnarray}
%
Making use of the specific properties of the various
operators involved, one can rewrite this as:
%
\begin{eqnarray}
  &\sim&
     - \mbox{Im}\,\mbox{Tr} [  (B_{xc}^i \sigma_z) (G_{01} \sigma_0)\, \xi_{SOC}^i (\sigma_z 
      \hat{l}_z) \nonumber \\
 && \times   G_{01} \sigma_0 (\sigma_x B_{xc}^j) \, (G_{01}
  \sigma_0) (\sigma_y B_{xc}^k) (G_{02} \sigma_z) ]^* \, , \nonumber
\end{eqnarray}
%
i.e.\ the original expression is recovered.
From this we may conclude that using 
Eq.\ (\ref{Eq_Free_Energy-TCI-1A})
within a more general fully relativistic framework
contributions occur which are also invariant under time reversal
and therefore  characterize a stationary
  state of the perturbed system.

\medskip

As a next step in the derivation of an expression
for the TCI interaction parameters,
the change in energy $\Delta {\cal E}$
in Eq.\ (\ref{Eq_Free_Energy-TCI-1A})
was mapped  to the
third term of the spin Hamiltonian given by
Eq.\ (\ref{Eq_Heisenberg_general}).\cite{MPE20}
Representing the Green function $ G_0$ by means of the 
multiple scattering formalism
$\Delta {\cal E}$ 
could be  expressed  by
a sum over products of the TCI parameters $J_{ijk}$
given  Eq.\ (\ref{Eq:J_XYZ})
and the corresponding scalar spin chirality 
$\hat{s}_i\cdot (\hat{s}_j \times  \hat{s}_k) $.
\medskip

As it was pointed out in  Section \ref{SEC:TCI},
the derivation of the  TCI term accounts automatically
for the symmetry properties of the scalar spin 
chirality $\hat{s}_i\cdot (\hat{s}_j \times \hat{s}_k)$.
This implies that multiplying the  TCI parameters by
$\hat{s}_i\cdot (\hat{s}_j \times \hat{s}_k)$ and
summing over all sites should  in particular be conform
with the behavior under time reversal of the initial expression,
i.e.\ the result  should be invariant under time reversal.
Following the discussions in  Section \ref{SEC:TCI-TOM} and in
Ref.\ [\onlinecite{MPE20}], the TCI term in the
spin Hamiltonian can be contracted to a form 
  accounting for only the counterclockwise contributions when summation
  over the sites: 
\begin{eqnarray}
  H^{(3)} &=& - \frac{1}{3!}\sum_{i,j,k:\circlearrowleft}
                   J_{\Delta} \hat{s}_i\cdot (\hat{s}_j \times \hat{s}_k) \; ,          
\label{Eq_Heisenberg_TCI-A}
\end{eqnarray}
with $J_{\Delta} = J_{ijk} - J_{ikj}$ giving the
non-zero contributions to the energy.
As a consequence of the mapping properties mentioned above, 
this parameter is antisymmetric with respect to time reversal
applied to the total system.
This important property was verified by electronic structure
calculations
with the exchange field changed in sign, that indeed led to
a flip in sign for the TCI parameters as well.
This property  ensures
that the energy 
given by Eq.\ (\ref{Eq_Heisenberg_TCI-A})
is time reversal invariant
as a consequence of the 
properties of the corresponding energy change given by Eq.\
(\ref{Eq_Free_Energy-TCI-1A}).
Connected with this,
it should be stressed once more that the expansion leading to Eq.\
(\ref{Eq_Free_Energy-TCI-1A}) is based on a -- in principle arbitrary --
reference state (see also the corresponding discussion in Ref.\
[\onlinecite{MPE20}]),
that addresses  all other parameters in the
spin Hamiltonian given by Eq.\ (\ref{Eq_Heisenberg_general}).
Accordingly, 
it represents the energy landscape in the vicinity of this state
in dependence on the actual spin configuration.
If the reference state is changed, e.g.\ by time reversal,
the expansion coefficients may change as well.

Finally, it is worth to compare with 
the energy mapping giving access to the DMI term,
where the standard  vector form of the  DMI parameters
ensures  corresponding scalar energy contributions.
In contrast to this situation,
the time reversal antisymmetry of the TCI parameters,
that nevertheless give finite
energy contributions invariant under time reversal,
is hidden.

\section{Computational details}
\label{SEC:Computational-scheme}

The first-principles exchange coupling parameters are calculated
using the spin-polarized relativistic KKR (SPR-KKR) Green
function method  \cite{SPR-KKR8.5,EKM11}.
The fully-relativistic mode was used except for the cases, where scaling
of the spin-orbit interaction was applied. 
All calculations have been performed using the atomic sphere
approximation (ASA), within the framework of the local 
spin density approximation (LSDA) to spin density 
functional theory (SDFT), using a parametrization for the exchange and
correlation potential as given by Vosko  et  al.\ \cite{VWN80}.
A cutoff $l_{max} = 2$ was used for the angular momentum 
expansion of the Green function.
Integration over the Brillouin zone (BZ) has been performed using a
$43 \times 43 \times 7$ k-mesh.

The calculations for 1ML of 3$d$ metals on a $M$(111) surface ($M$ = Ir,
Au) have been performed in the supercell geometry with (1ML Fe/3ML $M$)
layers separated by two vacuum layers. This decoupling was
sufficient in the present case to demonstrate the properties of the
exchange interaction parameters for the 2D system.  
The lattice parameter used were $a = 7.22$~a.u.\ for fcc Ir and $a =
7.68$ a.u. for fcc Au.

\begin{widetext}

\section{TOM} \label{SEC:TOM}
  
Eq. (\ref{Eq_Free_Energy-TCI-2}) can be modified by using the sum rule
for GF $\frac{dG}{dE} = -GG$ and integration by parts.

\begin{eqnarray}
\Delta {\cal E}^{(3)} &=& -\frac{1}{\pi} \mbox{Im}\,\mbox{Tr} \int^{E_F}
                    dE (E - E_F)\, [ VG_0 V G_0 \hat{\cal H}_B
                    \dot{G}_0 
                 +
                    V G_0 \hat{\cal H}_B G_0V \dot{G}_0 + \hat{\cal H}_B
                    G_0 V G_0 V \dot{G}_0 ] \;\nonumber\\
                &=& \frac{1}{\pi} \mbox{Im}\,\mbox{Tr}
                    \int^{E_F} dE V G_0 V G_0\hat{\cal H}_B G_0 
  +  \frac{1}{\pi} \mbox{Im}\,\mbox{Tr} 
                    \int^{E_F} dE V G_0 \hat{\cal H}_BG_0VG_0 
   + \frac{1}{\pi} \mbox{Im}\,\mbox{Tr}  
                    \int^{E_F} dE \hat{\cal H}_B
                    G_0 V G_0 V G_0 ]    \;\nonumber\\  
                && -\frac{1}{\pi} \mbox{Im}\,\mbox{Tr}
                    (E - E_F) V G_0 V G_0 \hat{\cal H}_B \dot{G}_0 |^{E_F} 
          +\frac{1}{\pi} \mbox{Im}\,\mbox{Tr} \int^{E_F} dE (E
                   - E_F)\, \frac{d}{dE}[ V G_0 V G_0 
                    \hat{\cal H}_B] G_0  \;\nonumber\\  
   && -\frac{1}{\pi} \mbox{Im}\,\mbox{Tr}
       (E - E_F) V G_0 V G_0 \hat{\cal H}_B \dot{G}_0 |^{E_F}
       +\frac{1}{\pi} \mbox{Im}\,\mbox{Tr} \int^{E_F}
                    dE (E - E_F)\, \frac{d}{dE}[ V G_0 \hat{\cal H}_B
      G_0V] G_0  \nonumber\\  
   &&  -\frac{1}{\pi} \mbox{Im}\,\mbox{Tr} 
       (E - E_F) V G_0 V G_0 \hat{\cal H}_B \dot{G}_0 |^{E_F} 
    +\frac{1}{\pi} \mbox{Im}\,\mbox{Tr} \int^{E_F}
                    dE (E - E_F)\, \frac{d}{dE}[ \hat{\cal H}_B
      G_0 V G_0 V] G_0  \;\nonumber\\
                &=& \frac{1}{\pi} \mbox{Im}\,\mbox{Tr}
                    \int^{E_F} dE V G_0 V G_0\hat{\cal H}_B G_0 +
                    \frac{1}{\pi} \mbox{Im}\,\mbox{Tr} 
                    \int^{E_F} dE V G_0 \hat{\cal H}_BG_0VG_0 +
                    \frac{1}{\pi} \mbox{Im}\,\mbox{Tr}  
                    \int^{E_F} dE \hat{\cal H}_B
                    G_0 V G_0 V G_0 ]    \;\nonumber\\    
                && +\frac{1}{\pi} \mbox{Im}\,\mbox{Tr} \int^{E_F} dE (E
                   - E_F)\, \frac{d}{dE}[ V G_0 V G_0 
                    \hat{\cal H}_B] G_0  \;\nonumber\\  
   && +\frac{1}{\pi} \mbox{Im}\,\mbox{Tr} \int^{E_F}
                    dE (E - E_F)\, \frac{d}{dE}[ V G_0 \hat{\cal H}_B G_0V] G_0  
    +\frac{1}{\pi} \mbox{Im}\,\mbox{Tr} \int^{E_F}
                    dE (E - E_F)\, \frac{d}{dE}[ \hat{\cal H}_B
                    G_0 V G_0 V] G_0  \;
\label{Eq_Free_Energy-TCI-3}
\end{eqnarray}

After partial integration, the terms without involving an integral
should vanish due to the factor $(E - E_F)|^{E_F}$. 
Taking the energy derivatives in the last three integrals, we obtain
\begin{eqnarray}
&&-\frac{1}{\pi} \mbox{Im}\,\mbox{Tr} \int^{E_F}
                    dE (E - E_F)\, [ V G_0 V G_0 \hat{\cal H}_B
                    \dot{G}_0 +
                    V G_0 \hat{\cal H}_B G_0V \dot{G}_0 + \hat{\cal H}_B
                    G_0 V G_0 V \dot{G}_0 ] \;\nonumber\\
                &=& \frac{1}{\pi} \mbox{Im}\,\mbox{Tr}
                    \int^{E_F} dE\, V G_0 V G_0\hat{\cal H}_B G_0 +
                    \frac{1}{\pi} \mbox{Im}\,\mbox{Tr} 
                    \int^{E_F} dE\, V G_0 \hat{\cal H}_BG_0VG_0 +
                    \frac{1}{\pi} \mbox{Im}\,\mbox{Tr}  
                    \int^{E_F} dE\, \hat{\cal H}_B
                    G_0 V G_0 V G_0     \;\nonumber\\    
                && +\frac{1}{\pi} \mbox{Im}\,\mbox{Tr} \int^{E_F} dE (E
                   - E_F)\, [ V \dot{G}_0 V G_0 
                    \hat{\cal H}_B] G_0 + \frac{1}{\pi} \mbox{Im}\,\mbox{Tr} \int^{E_F} dE (E
                   - E_F)\, [ V G_0 V \dot{G}_0 
                    \hat{\cal H}_B] G_0  \;\nonumber\\  
   && +\frac{1}{\pi} \mbox{Im}\,\mbox{Tr} \int^{E_F}
                    dE (E - E_F)\, [ V \dot{G}_0 \hat{\cal H}_B G_0V] G_0 + \frac{1}{\pi} \mbox{Im}\,\mbox{Tr} \int^{E_F}
                    dE (E - E_F)\, [ V G_0 \hat{\cal H}_B \dot{G}_0V] G_0  \nonumber\\  
   && 
    +\frac{1}{\pi} \mbox{Im}\,\mbox{Tr} \int^{E_F}
                    dE (E - E_F)\, [ \hat{\cal H}_B
                    \dot{G}_0 V G_0 V] G_0  + \frac{1}{\pi} \mbox{Im}\,\mbox{Tr} \int^{E_F}
                    dE (E - E_F)\, [ \hat{\cal H}_B
                    G_0 V \dot{G}_0 V] G_0  \;.
\label{Eq_Free_Energy-TCI-3a}
\end{eqnarray}

\end{widetext}

Using the invariance of the trace of matrix product w.r.t. cyclic
permutations, one can combine the latter integrals and bring them to the
left. With this one arrives at the expression
\begin{eqnarray}
\Delta {\cal E}^{(3)} &=&  \frac{1}{3} \frac{1}{\pi} [ \mbox{Im}\,\mbox{Tr}
                    \int^{E_F} dE V G_0 V G_0\hat{\cal H}_B G_0
                    \nonumber \\
  &&+
                    \mbox{Im}\,\mbox{Tr} 
     \int^{E_F} dE V G_0 \hat{\cal H}_BG_0VG_0 
                    \nonumber \\
  &&  +
                    \mbox{Im}\,\mbox{Tr}  
                    \int^{E_F} dE \hat{\cal H}_B
                    G_0 V G_0 V G_0 ]    \;.
\label{Eq_Free_Energy-TCI-4}
\end{eqnarray}
Representing the Green functions in terms of multiple-scattering
formalism, this expression can be reduced to the expression
\begin{eqnarray}
\Delta {\cal E}^{(3)} &=&
 \frac{1}{3}\frac{1}{\pi} \sum_{i \neq j\neq k}\mbox{Im}\,\mbox{Tr} \int^{E_F}
                          dE\,  \nonumber \\
                      && \times \bigg[                         
                         \langle Z_i|\hat{\cal H}_B| Z_i \rangle \tau_{ij}
                         \langle Z_j| \delta v_{j} |Z_j\rangle \tau_{jk}  
                         \langle Z_k| \delta v_{k} |Z_k \rangle \tau_{ki}
                           \nonumber \\  
                      &&                        
                        + \langle Z_i|\delta v_{i} |Z_i \rangle \tau_{ij}
                         \langle Z_j|\hat{\cal H}_B|Z_j\rangle \tau_{jk} 
                         \langle Z_k| \delta v_{k} |Z_k \rangle \tau_{ki}
                           \nonumber \\  
                      &&                        
                        + \langle Z_i|\delta v_{i} | Z_i \rangle \tau_{ij}
                         \langle Z_j| \delta v_{j} |Z_j\rangle \tau_{jk}  
                         \langle Z_k|\hat{\cal H}_B| Z_k \rangle \tau_{ki} \bigg] \;,
               \nonumber \\       
\label{Eq_Free_Energy-TCI-5}
\end{eqnarray}
which has a similar form as the expression for the energy associated with
the three-spin chiral interactions given previously\cite{MPE20}.
Here the perturbation $\delta v_{i}$ in the system is associated with the
non-coplanar magnetic texture. Following the idea used to derive the
expression for the TCI\cite{MPE20}, we create a $2q$ spin modulation
according to
\begin{eqnarray}
  \hat{s}_i &=&
 (\sin(\vec{q}_1 \cdot \vec{R}_i) \, \cos (\vec{q}_2 \cdot \vec{R}_i), \; 
                \sin(\vec{q}_2 \cdot \vec{R}_i), \nonumber \\
&&  \cos (\vec{q}_1 \cdot
 \vec{R}_i) \, \cos (\vec{q}_2 \cdot \vec{R}_i) ) \;,
\label{spiral2}
\end {eqnarray}
which is characterized by two wave vectors, $\vec{q}_1$ and $\vec{q}_2$,
orthogonal to each other. 
Taking the second-order derivatives with respect to $\vec{q}_1$ and
$\vec{q}_2$ in the limit $q_1 \to 0$, $q_2 \to 0$, Eq.\
(\ref{Eq_Free_Energy-TCI-5}) can be reduced to the form
\begin{eqnarray}
 \Delta {\cal E}^{(3)}  &=&  \frac{1}{3!} \sum_{i \neq j\neq k} \hat{s}_i\cdot (\hat{s}_j \times \hat{s}_k)
                            \frac {1}{4\pi} \mbox{Im}\, \mbox{Tr}
                            \int^{E_F} dE \nonumber \\ 
 &&                    
\times \Big[ \underline{T}^{i, x}\, \underline{\tau}^{ij}
\underline{T}^{j, y}\, \underline{\tau}^{jk}
               \underline{H}_B^{k}\,  \underline{\tau}^{ki} 
- \underline{T}^{i, y}\, \underline{\tau}^{ij}
 \underline{T}^{j, x}\, \underline{\tau}^{jk}
     \underline{H}_B^{k}\,  \underline{\tau}^{ki}
               \nonumber \\       
&&     - \underline{T}^{i, x}\, \underline{\tau}^{ij}
\underline{H}_B^{j}\, \underline{\tau}^{jk}
               \underline{T}^{k, y}\, \underline{\tau}^{ki}\;
+ \underline{T}^{i, y}\, \underline{\tau}^{ij}
\underline{H}_B^{j}\, \underline{\tau}^{jk}
               \underline{T}^{k, x}\, \underline{\tau}^{ki} \;
               \nonumber \\       
&& +
    \underline{H}_B^{i} \, \underline{\tau}^{ij} 
\underline{T}^{j, x}\, \underline{\tau}^{jk}
               \underline{T}^{k, y}\, \underline{\tau}^{ki}\;
 - \underline{H}_B^{i}\,  \underline{\tau}^{ij}
\underline{T}^{j, y}\,\underline{\tau}^{jk}
               \underline{T}^{k, x}\, \underline{\tau}^{ki} \Big] \;.
               \nonumber \\       
\label{Eq_Free_Energy-TCI-7} 
\end{eqnarray}
That in turn, taking into account $\hat{\cal H}_B = - \hat{\vec l} \cdot
\vec{B}$, leads to the topological orbital moment 
\begin{eqnarray}
   L^{\rm{TO}}  &=& \frac{1}{3!} \sum_{i \neq j\neq k}
                   \chi^{\rm{TO}}_{ijk} \hat{s}_i\cdot (\hat{s}_j \times \hat{s}_k)                               
                               \nonumber \\
 &=& - \frac{1}{3!}  \sum_{i \neq j\neq k}  \hat{s}_i\cdot (\hat{s}_j \times \hat{s}_k) \frac {1}{4\pi} \mbox{Im}\, \mbox{Tr} \int^{E_F} dE \nonumber \\
 &&                    
\times \Big[ \underline{T}^{i, x}\, \underline{\tau}^{ij}
\underline{T}^{j, y}\, \underline{\tau}^{jk}
               \underline{l}_z^{k}\,  \underline{\tau}^{ki} 
- \underline{T}^{i, y}\,  \underline{\tau}^{ij}
 \underline{T}^{j, x}\, \underline{\tau}^{jk}
     \underline{l}_z^{k}\,  \underline{\tau}^{ki}
               \nonumber \\       
&&     - \underline{T}^{i, x}\, \underline{\tau}^{ij}
\underline{l}_z^{j}\, \underline{\tau}^{jk}
               \underline{T}^{k, y}\, \underline{\tau}^{ki}\;
+ \underline{T}^{i, y}\, \underline{\tau}^{ij}
\underline{l}_z^{j}\, \underline{\tau}^{jk}
               \underline{T}^{k, x}\, \underline{\tau}^{ki} \;
               \nonumber \\       
&& +
    \underline{l}_z^{i} \, \underline{\tau}^{ij} 
\underline{T}^{j, x}\, \underline{\tau}^{jk}
               \underline{T}^{k, y}\, \underline{\tau}^{ki}\;
 - \underline{l}_z^{i}\,  \underline{\tau}^{ij}
\underline{T}^{j, y}\,\underline{\tau}^{jk}
               \underline{T}^{k, x}\, \underline{\tau}^{ki} \Big] \;.
               \nonumber \\       
\label{Eq:TOM} 
\end{eqnarray}


%

\end{document}